\documentstyle[sg,psfig
]{mn}

\renewcommand{\etal} {et al.}                
\renewcommand{\eg}   {e.g.}                  
\renewcommand{\ie}   {i.e.}                  

\newcommand{\arcsecintxtE} {\arcsecE}   
\newcommand{\arcminintxtE} {\arcminE}   

\title[\HI\ Observations of Interacting Galaxy Pair NGC~4038/9]
      {\HI\ Observations of Interacting Galaxy Pair NGC~4038/9}

\author[Gordon \etal]
{ Scott~Gordon$^1$\thanks{E-mail: gordon@physics.uq.edu.au},
  B\"arbel~Koribalski$^2$,
  Keith~Jones$^1$ \\
$^1$Department of Physics, University of Queensland, St. Lucia,
    Brisbane, QLD 4072, Australia \\
$^2$Australia Telescope National Facility, CSIRO,
    P.O. Box 76, Epping, NSW 1710, Australia \\
}

\date{Received date; accepted date}

\begin{document}

\maketitle
						  
\begin{abstract}

We present the results of new radio interferometer \HI\ line observations for
the merging galaxy pair NGC~4038/9 (`The Antennae'), obtained using the
Australia Telescope Compact Array.
The results improve substantially on those of van der Hulst (1979) and show
in detail the two merging galactic disks and the two tidal tails produced by
their interaction. The small edge-on spiral dwarf galaxy ESO~572--G045 is also
seen near the tip of the southern tail, but distinct from it. It shows no signs
of tidal interaction. The northern tidal tail of the Antennae shows no \HI\ 
connection to the disks and has an extension towards the west. The southern
tidal tail is continuous, with a prominent \HI\ concentration at its tip,
roughly at the location of the tidal dwarf galaxy observed optically by
Mirabel, Dottori \& Lutz (1992). Clear velocity structure is seen along the
tidal tails and in the galactic disks. Radio continuum images at 20\,\cm\ 
and 13\,\cm\ are also presented, showing the disks in detail.
\end{abstract}


\section{Introduction}
\label{section:Introduction}
\indent

\begin{figure*} 
\begin{center}
\mbox{\psfig{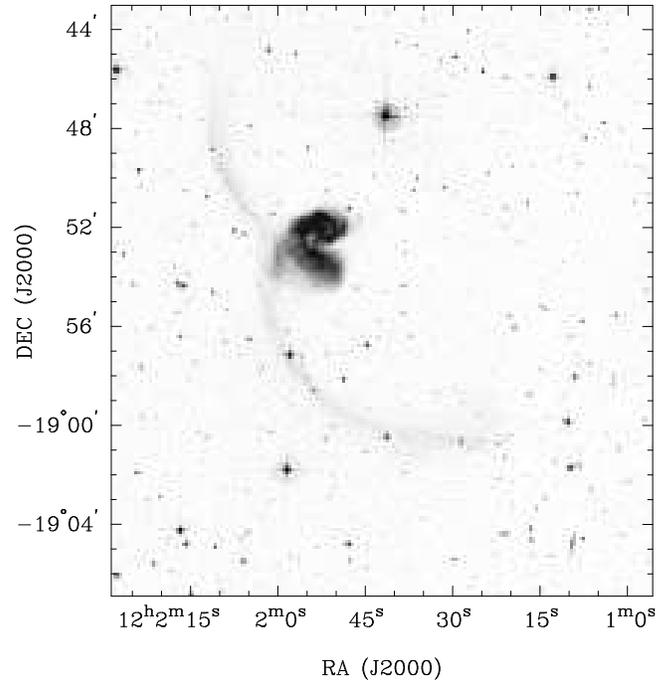}}
\end{center}
\caption[Optical Image of the Antennae]{
\label{figure:AntennaeDSS}
Digitised Sky Survey (DSS) optical image of the Antennae (NGC~4038/9). }
\end{figure*}

The Antennae (NGC~4038/4039 = Arp~244, see Fig.~\ref{figure:AntennaeDSS}) are
a pair of closely-interacting disk galaxies in the advanced stages of merging,
named after the appearance of the two spectacular galactic tails produced by
gravitational tidal forces during the interaction. There are a number of
similar systems known, including `The Mice' (NGC~4676A/B) described by Hibbard
(1995) and `The Superantennae' (IRAS~19254--7245) described by Mirabel \etal\
(1991). Also visible to the south-west of the southern tail of the Antennae is
the edge-on spiral dwarf galaxy ESO~572--G045 (Fig.~\ref{figure:AntennaeDSS}),
which has no measured optical velocity. Some parameters of the Antennae and
ESO~572--G045 are given in Table~\ref{table:TheGroup}.

On a large scale, the Antennae appears roughly symmetrical, although the
southern tail is more prominent both optically (Fig.~\ref{figure:AntennaeDSS})
and in \HI\ (van der Hulst 1979; this paper). The two galaxies retain distinct
nuclei as seen from optical and CO observations. However, the galactic disks,
while still present, are colliding and have severely disrupted one another. 
The two nuclei are currently $\sim$1\farcm13 or $\sim$6.3~kpc apart 
across the line-of-sight, according to CO measurements by Stanford \etal\ 
(1990). As a result of the collision, vigorous star formation is occurring in
the disks and in the overlap region between them. This is evidenced by a 
variety of data, for example HST optical observations (Whitmore \& Schweizer
1995; Whitmore \etal\ 1999), and ISO infrared observations 
(\eg\ Vigroux \etal\ 1996) which show a high infrared luminosity in the two
nuclei and particularly in the overlap region.

The tidal tails, disturbed structure, and star formation, are all classic
features of interacting galaxies. Given this and the closeness of the system
($D = 19.2\,\MpcE$\footnote{This paper assumes a Hubble constant of \Ho{75}
and a velocity relative to the Local Group of $1439~\kmpsE$ for the Antennae
(Whitmore \& Schweizer 1995 from de Vaucouleurs \etal\ 1991) giving a Hubble
distance of 19.2\,\Mpc\ and a linear scale of 1\arcmin\ $\approx 5.6$~kpc.}),
the Antennae have attracted a lot of attention in observational and theoretical
studies of galaxy interaction.

\subsection{History of the Antennae}
\label{section:IntroHistory}
\indent

The Antennae were first seen by C.O. Lampland of Lowell Observatory in 1917
(Parker 1990), and were later described by Shapley \& Paraskevopoulos (1940),
and Minkowski (1957), who identified the pair with a radio source and first
described them as interacting galaxies. They were also noted as number 244 
in the Arp atlas of peculiar galaxies (Arp 1966).

Systems resembling the Antennae were produced in the pioneering computer
modelling simulations by Toomre \& Toomre (1972), providing the first
theoretical proof that features such as galactic tails were produced by
gravitational tides between interacting galaxies, leading to much of the
present understanding of their evolution. Clutton-Brock (1972) published work
similar to the Toomres', but less extensive. The Antennae are the first or
`earliest' of the 11 systems in the `Toomre Sequence' (Toomre 1977), and are
still used for modelling (\eg\ Barnes 1988; Mihos, Bothun \& Richstone 1993).
Collectively, these simulations have disproved, for most interacting systems,
earlier explanations for galactic pairs and interaction effects, such as the
suggestion by Arp (1969) that when two galaxies of different sizes were
interacting, the smaller one was an ejected component of the larger.

\subsection{Previous Observations}
\label{section:IntroPreviousObs}
\indent

Observations of the Antennae have been made at many radio, infrared, optical,
and X-ray wavelengths, including many emission lines such as \Halpha\ (Amram
\etal\ 1992), \CII\ (Nikola \etal\ 1998), CO~J=1--0 (Sanders \& Mirabel 1985;
Stanford \etal\ 1990; Young \etal\ 1995; Gao \etal\ 1998; Wilson \etal\ 2000;
Gao \etal\ 2001), and \HI\ (Huchtmeier \& Bohnenstengel 1975; van der Hulst
1979), \Halpha, \Hbeta, \OI, \OIII, \HeI\ and \SII\ (Rubin, Ford \& D'Odorico
1970), and also radio continuum (Condon 1983; Condon 1987; Hummel \& van der
Hulst 1986).

More recently, radio observations of the Antennae have been made using the VLA.
These include high-resolution 6\,\cm\ and 3.5\,\cm\ observations by Neff \&
Ulvestad (2000), and \HI\ observations by Mahoney, Burke \& van der Hulst
(1987) and by Hibbard, van der Hulst \& Barnes (2001, in prep).

Infrared continuum studies have been made using the ISO satellite by Fischer
\etal\ (1996), Kunze \etal\ (1996), Vigroux \etal (1996), and Mirabel \etal\
(1998). X-ray studies have been made using the Einstein Observatory by Fabbiano
\& Trinchieri (1983), ROSAT by Read \etal\ (1995) and Fabbiano \etal\ (1997),
ASCA by Sansom \etal\ (1996), and most recently using Chandra by Fabbiano,
Zezas \& Murray (2000).

\subsection{The Inner Disks}
\label{section:IntroInnerDisks}
\indent

There are numerous \HII\ regions distributed in a ring in the NGC~4038 
(northern) disk and a few also along an arc in the NGC~4039 (southern) disk.
The shape of their distribution is roughly that of a letter `e', and is seen
clearly as knots, or compact objects, in blue optical images as well as in
\Halpha\ (Amram \etal\ 1992) and in the radio continuum (Hummel \& van der
Hulst 1986). The Hubble Space Telescope resolved the \HII\ regions into many
groups and clusters of young stars, with over 700 mostly blue compact objects
20--30~\pc\ wide (Whitmore \& Schweizer 1995; Whitmore \etal\ 1999). 

The location of one particular group of optical and radio continuum knots in
the overlap region between the two galactic disks, combined with the youth
of the stars, suggests that this star formation activity is caused by the 
on-going interaction. Further high-resolution mid-infrared studies using the
ISO satellite (Mirabel \etal\ 1998) show the strongest star formation in the
overlap region, accounting for half of the $15\,\umE$ flux, rather than in the
more optically spectacular star-forming ring. This is explained as a result of
obscuration by the large quantity of dust, which tends to hide star-forming
regions at optical wavelengths but also makes them highly visible in the 
infrared (Kunze \etal\ 1996; Mirabel \etal\ 1998; Wilson \etal\ 2000). Wilson
\etal\ (2000) produced detailed CO maps and molecular mass estimates for the
disk region using the Caltech mm-array. The maps show seven giant molecular
cloud complexes, including the two nuclei and five complexes in the disk 
overlap region which have on-going massive star formation. Gao \etal\ (2001) 
made a similar study combining single-dish and interferometric observations,
leading to an estimated molecular gas mass of $\baseexpE{1.38}{10}\,\MsolarE$ 
(adopting $D=19.2\,\MpcE$), implying a modest star formation efficiency.

\subsection{Southern-Tail Tidal Dwarf Galaxy}
\label{section:IntroTailDwarf}
\indent
 
At the tip of the southern tail is an object thought to be a dwarf galaxy
formed from material ejected from the galactic disks during the interaction
(Mirabel, Dottori \& Lutz 1992). It contains many hot, young stars, and
coincides with a major concentration of \HI\ which has been noted in earlier
studies (van der Hulst 1979).

This dwarf is of particular interest because it matches the results of computer
modelling, confirming that tidally-disturbed material can become concentrated
enough to form bound objects (Barnes \& Hernquist 1992, 1996). Additionally,
observing a dwarf galaxy soon after its formation could shed light on the
origin and star-forming properties of such galaxies, about which little is
currently known.

\vspace{3mm}\noindent \rule{50mm}{0.25mm} \\

Section~2 describes the observations and data reduction. This is followed by 
a summary of the major features in Section~3. The implications of the data,
and its relationship to previous work, is given in Section~4. Finally, 
Section~5 presents a summary of our new and interesting results.

\section{Observations \& Data Reduction}
\label{section:Observations}
\indent

We have observed the Antennae group using the Australia Telescope Compact
Array (ATCA\footnote{The Australia Telescope Compact Array is part of the
  Australia Telescope which is funded by the Commonwealth of Australia for
  operation as a National Facility managed by CSIRO.}),
a linear east-west oriented radio interferometer with six 22-m parabolic 
dishes.  Table~\ref{table:ObsDetails} gives details of the observations. In
total, four successful 12~hour observations were obtained (plus one additional
2--3~hour run) using different antenna arrays chosen to give baselines spread
across as wide a range as possible. The dual 20/13\,\cm\ feedhorns on the 
ATCA, producing two independent IFs, were used to observe in both bands 
simultaneously. Data reduction was performed, unless otherwise stated, using
the MIRIAD analysis package (Sault, Teuben \& Wright 1995). All radio images
presented have been corrected for primary-beam attenuation. In some figures,
pixels below $3\sigma$ have been masked out prior to correction to exclude
noise. All velocities are in the heliocentric reference frame and the
`optical' definition $v=c\left(\Delta\lambda/\lambda_{\circ}\right)$. The
measured \HI\ velocities have an uncertainty of about 5\,\kmps. Positions 
are expressed in the J2000 system, including any from other papers.

\subsection{\HI\ Line and 20-cm Continuum Observations}
\label{section:Obs20cm}
\indent

The 20\,\cm\ data (IF1; see Table~\ref{table:ObsDetails}) covered the
red-shifted \HI\ emission line of neutral atomic hydrogen, plus some continuum 
channels to each side. The data was Hanning-smoothed to a resolution of 
$\sim6.6\,\kmpsE$. Channels at each edge of the passband (attenuated by 
50\% or more) were removed, giving a remaining bandwidth of $\sim6.8\,\MHzE$. 
This band was used to produce separate \HI\ line and 20-cm continuum datasets,
using a first-order polynomial (or baseline) fit to the line-free channels 
to each side of the spectrum.

The 20\,\cm\ (1413\,\MHz) flux for PKS~1934--638 is $\sim$14.88~Jy.
The derived antenna gains ranged from 0.99 to 1.06, indicating a typical
$\sim$3--4\% change during calibration.

The line data for the individual observations were then combined and
Fourier-transformed to produce a spectral image cube. The visibilities were
weighted using `robust' weighting with the robustness parameter equal to 1,
chosen to give a good compromise between sensitivity and resolution. This 
image cube was cleaned by deconvolution using a combined H\"ogbom/Clark/Steer
algorithm (H\"ogbom 1974, Clark 1980, Steer \etal\ 1984). The \HI\ maps have
a channel width of $10\,\kmpsE$; the pixel size is 5\arcsec\ $\times$ 5\arcsec.
The r.m.s. residual after cleaning is $\sim1~\mJypbE$. The map was restored
using a circular restoring beam of 42\arcsec.

The \package{AIPS} task \task{MOMNT} was used to produce maps of \HI\ 
integrated intensity (moment-0), intensity-weighted mean velocity (moment-1)
and velocity-dispersion (moment-2). To smooth the cube spatially and in 
velocity we used a three pixel `boxcar' function and a five channel `Hanning'
function, respectively. A lower flux cutoff of $2.1~\mJypbE$ ($\sim3\sigma$ 
after smoothing) was applied.

The 20\,\cm\ continuum map (Fig.~\ref{figure:Antennae20_13CM_Overlay}a) was
made using `uniform' weighting resulting in a noise after cleaning of
0.23\,\mJypb, and a $16\arcsecintxtE$ restoring beam was used.

\subsection{13-cm Continuum Observations}
\label{section:Obs13cm}
\indent

The 13\,\cm\ data (IF2; see Table~\ref{table:ObsDetails}) had 128\,\MHz\
bandwidth and full polarisation information. After excluding edge channels, 
the bandwidth was 100\,\MHz. No polarisation was observed above $1\%$.

The 13\,\cm\ continuum map (Fig.~\ref{figure:Antennae20_13CM_Overlay}b) was
made using `uniform' weighting, resulting in a noise after cleaning of
0.056\,\mJypb\ using a $9.5\arcsecintxtE$ restoring beam.

The flux of PKS~1934--638 ranges from $\sim$11.5~Jy 2378\,\MHz\ to
$\sim$11.1~Jy at 2496\,\MHz. The derived antenna gains ranged from 0.99 to
1.05, giving a typical change during calibration of $\sim3\%$.

\section{Results}
\label{section:Results}

\subsection{\HI\ Channel Map Summary}
\label{section:ResultsHIChannelMaps}
\indent

Neutral hydrogen in the Antennae group is observed in the velocity range 
$\sim1340-1946\,\kmpsE$ (see Fig.~\ref{figure:HIChannelMaps}). At low 
velocities the \HI\ emission consists of a single peak in the disk overlap
region, generally brightening with increasing velocity. Above 1560\,\kmps\
a second peak to the south is seen, moving southward and forming the inner 
half of the southern tail. A third peak beginning at $\sim1610\,\kmpsE$ 
forms the outer half of that tail, the two meeting near the tail's center
and disappearing at $\sim1780\,\kmpsE$. The northern tail is also detected
($\sim1540-1620~\kmpsE$) and possesses a previously unknown extension towards
the west at a similar velocity. 

The dwarf galaxy ESO572--G045 about 30\,\kpc\ south-west of the tip of the
southern tail is detected in the velocity range $\sim1650-1750\,\kmpsE$.

\subsection{\HI\ Statistics}
\label{section:ResultsHIStatistics}
\indent

The disk region and the southern tail have similar \HI\ masses of
$\sim\baseexpE{2.3}{9}\,\MsolarE$ and $\sim\baseexpE{2.5}{9}\,\MsolarE$,
respectively (see Table~\ref{table:HIstats}). The exact division between
the two regions is arbitrary since there is no clear boundary between them,
although the southern tail only has velocities above $\sim1590\,\kmpsE$.
The northern tail has an \HI\ mass of $\sim\baseexpE{0.2}{9}\,\MsolarE$.

We calculate an \HI\ mean velocity for the Antennae, of $\sim1664\,\kmpsE$,
obtained from the moment maps (Fig.~\ref{figure:HIMomentMaps}~a,b) (the mean
for the disks is $\sim1620\,\kmpsE$). This is similar to the $\sim1660\,\kmpsE$
systemic velocity obtained from optical measurements of the nuclear velocities
(see Table~\ref{table:TheGroup}).

The column density in Table~\ref{table:HIstats} is obtained from the \HI\ 
moment-0 map using the following formula:
\begin{equation}\label{eqn:HI_ColumnDensity}
N_{\HI} = \baseexpE{1.11}{24}\,\psqcmE \frac{\int{B(v)dv}}
      {\JypbkmpsE} \frac{{\rm arcsec}^2} {\theta_1\theta_2}~~~,
\end{equation}
where $B(v)$ is surface brightness, $v$ is velocity, and $\theta_1$ and 
$\theta_2$ are the major and minor beam FWHM. For our \HI\ maps with a
resolution of 42\arcsec\ there is $\baseexpE{6.23}{20}\,\psqcmE$
per \Jypbkmps\ of moment-0.

The \HI\ mass is calculated from the integrated \HI\ flux, $F(v)$, using the
following equation (Giovanelli \& Haynes 1988; Roberts 1975):

\begin{equation}\label{eqn:HI_Mass}
M_{HI} = \baseexpE{2.356}{5}\,\MsolarE {\left(\frac{D}{\MpcE}\right)}^2
         \frac{\int{F(v)dv}}{\JykmpsE}~~~.
\end{equation}
For $D=19.2\,\MpcE$ this gives \baseexp{8.68}{7}\,\Msolar\ per \Jykmps\
of \HI\ flux.

\subsection{The Galactic Disks}
\label{section:ResultsGalacticDisks}
\indent

\begin{figure*} 
\begin{tabular}{cc}
\mbox{\psfig{file=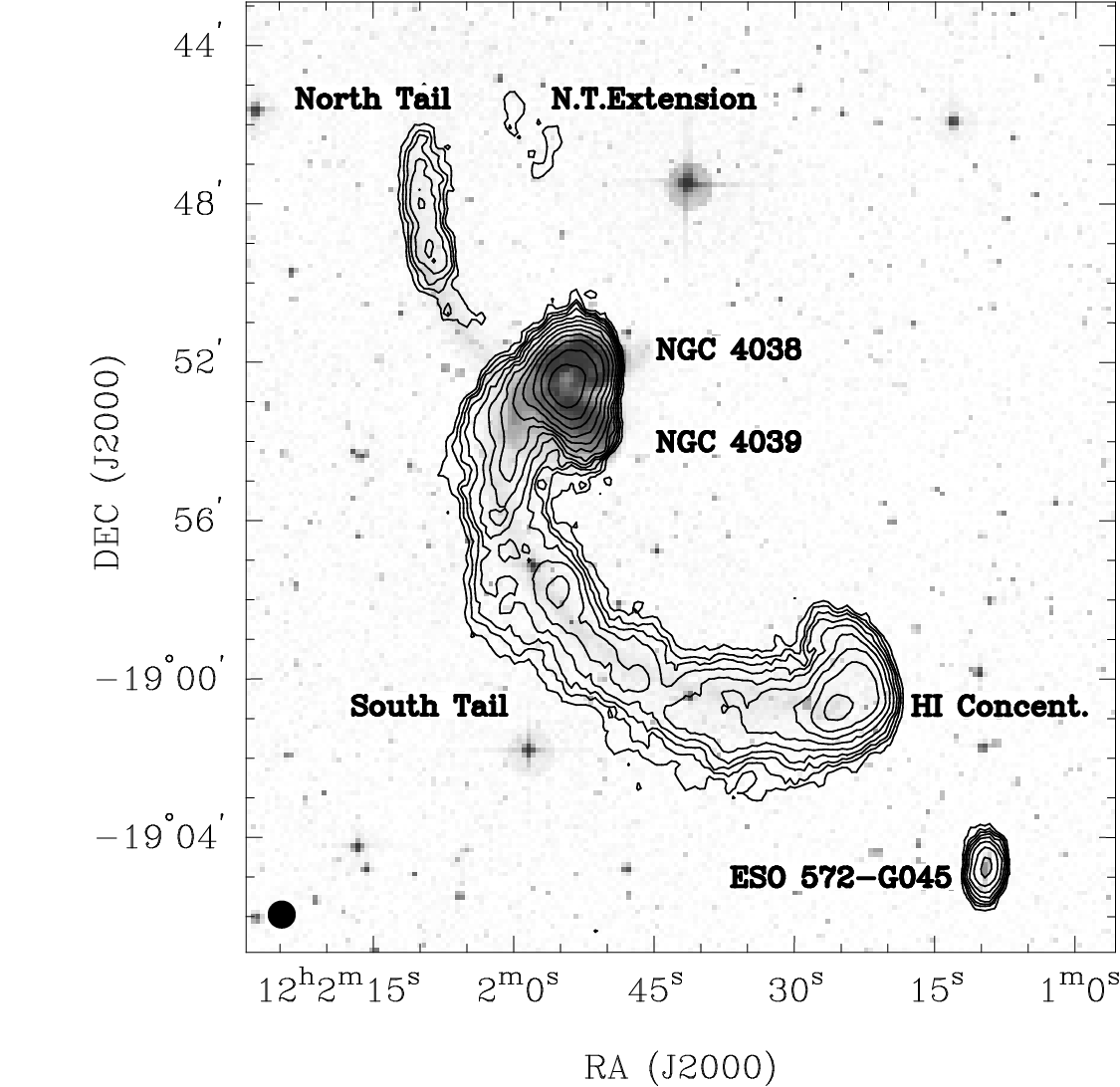,width=8.4cm}} &
\mbox{\psfig{file=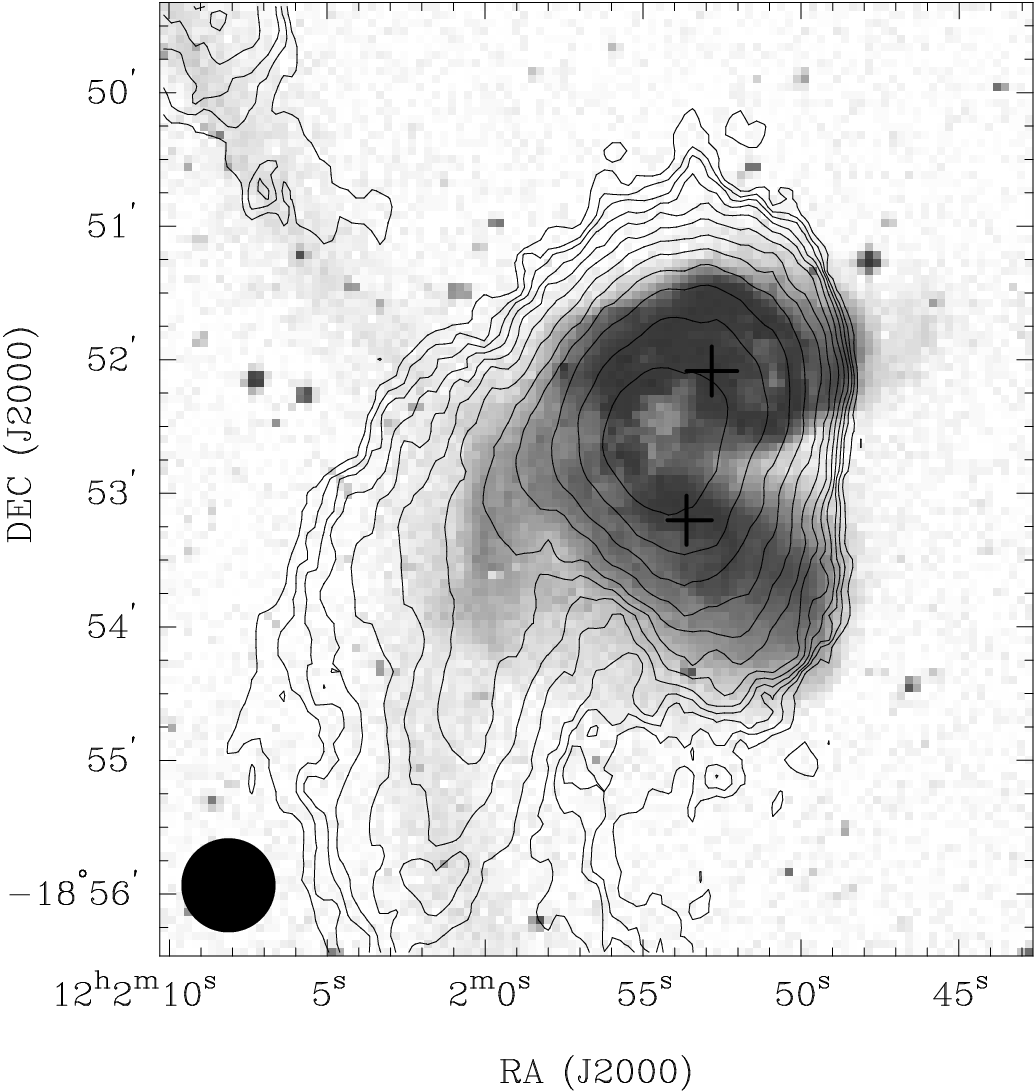,width=8.4cm}}
\\
{\bf(a)} \HI\ moment-0 (integrated intensity) &
{\bf(b)} \HI\ moment-0 (a closeup of {\bf(a)}) \vspace{5mm} \\
\mbox{\psfig{file=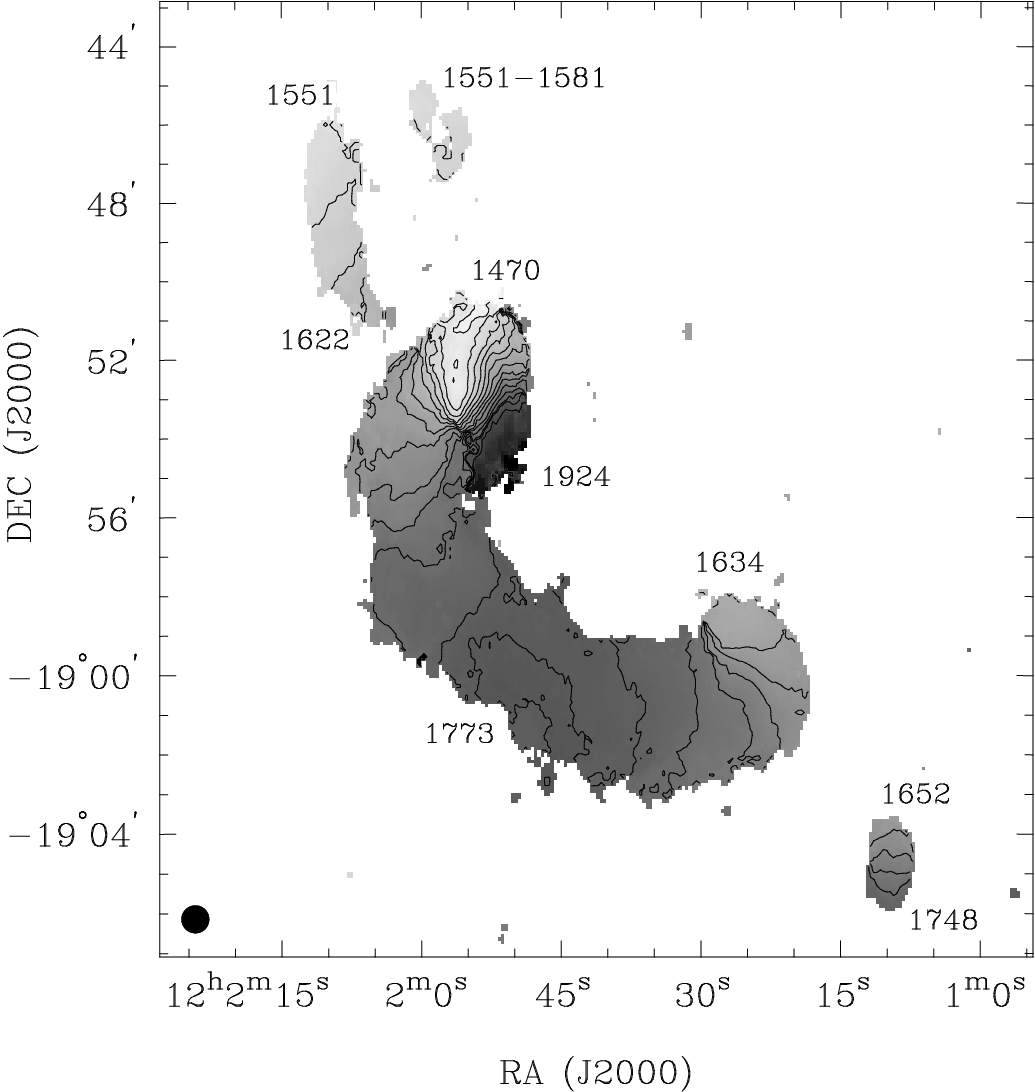,width=8.4cm}} &
\mbox{\psfig{file=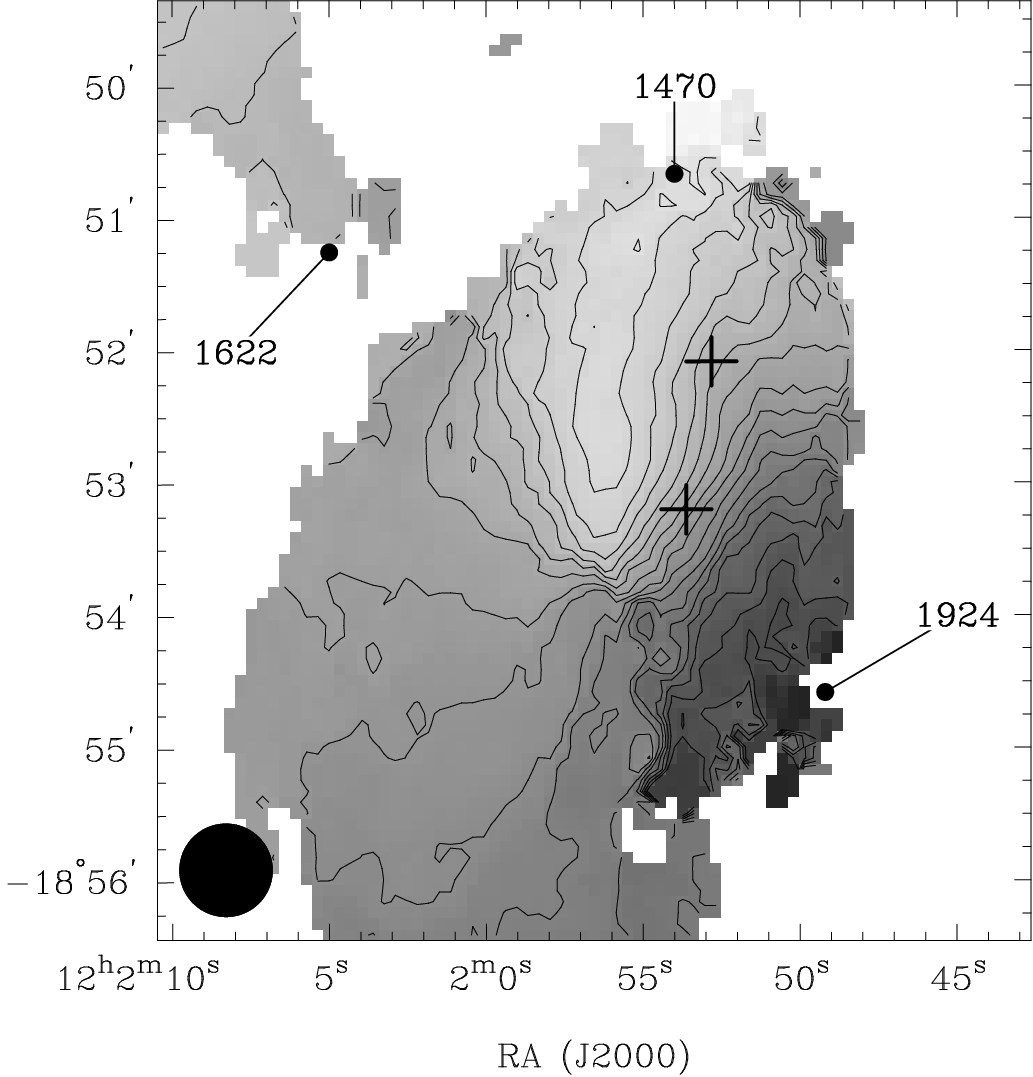,width=8.4cm}}
\\
{\bf(c)} \HI\ moment-1 (mean velocity) &
{\bf(d)} \HI\ moment-1 (a closeup of {\bf(c)}) \vspace{5mm} \\
\end{tabular}
\caption[\HI\ Moment Maps for The Antennae]{
\label{figure:HIMomentMaps}
\HI\ moment maps: those to the right are close-ups, enlarged by a factor of
$\sim$3, and with crosses marking the galactic nuclei (Stanford \etal\ 1990).
~~~{\bf(a,\,b)} The \HI\ distribution (moment-0 in contours) overlaid on the
optical image (see Fig.~\ref{figure:AntennaeDSS}). Contours: 60, 120, 170, 240,
340, 480, 680, 960, 1360, 1920, 2715, 3840, 5430~\Jypbmps, \ie\ powers of
$2^{1/2}$. This is also $\sim\baseexpE{6.66}{22}\times2^{n/2}~\psqcmE$
($n\geq0$). ~~~{\bf(c,\,d)} The \HI\ mean velocity field (moment-1). Contours:
1520--1780~\kmps, in steps of 20~\kmps, plus 1800, 1820, and 1840\,\kmps\ in the
close-up. ~~~{\bf(e,\,f)} The \HI\ velocity dispersion (moment-2). Contours: 15,
20, 25\,\kmps, and 30--140~\kmps, in steps of 10\,\kmps. Labels indicate
individual components as well as velocity peaks and ranges. The restoring beam
(42\arcsec) is shown at bottom left corner.  }
\end{figure*}

The overall shape of the \HI\ in the disk region and at the base of the 
southern tail (Fig.~\ref{figure:HIMomentMaps}b) follows the shape of the 
optical emission, except that: 
{\bf(1)} the \HI\ spans the gap between the western ends of the two disks
 and ends well inside the optical image, particularly in the northern disk,
 whereas galaxies usually have more extended \HI; 
{\bf(2)} the northern (and smaller) tidal tail is not connected to the disks
 in \HI, with a $\sim1\arcminintxtE$ gap in the \HI\ distribution;
and {\bf(3)} the \HI\ intensity peak is midway between the optical nuclei. 

The peak in the \HI\ distribution lies about one third of the distance
between the two nuclei, about $32\arcsecintxtE$ (3\,\kpc) north of the
center of NGC~4039 (Table~\ref{table:HIstats}).

\HI\ spectra taken throughout the disks generally show a single peak in 
the spectrum. However, there are two velocity peaks in a wide region midway
between the centres of the two disks (Fig.~\ref{figure:HINuclearSpectra}),
at $\sim1542$ and $\sim1630~\kmpsE$. These peaks are also visible in the
overall disk spectrum (Fig.~\ref{figure:HISpectra}a), but they don't match
the systemic velocities of the disks, which are $\sim1663$ and 
$\sim1655\,\kmpsE$ (Table~\ref{table:TheGroup}).

The velocities in the disks (Fig.~\ref{figure:HIMomentMaps}c,d) range from
1470\,\kmps\ at the northern disk edge, to 1920\,\kmps\ close to the
south-western disk edge. 

The disks have a substantial \HI\ velocity dispersion (see Fig.~2f), with 
a peak of $\sim150\,\kmpsE$ about $30\arcsecintxtE$ south of the NGC~4039 
nucleus (Table~\ref{table:TheGroup}), far from the intensity peak. The high 
dispersion results from the dual-peaked spectrum in its vicinity.

The dominant feature in the south-east is the base of the southern tail,
which attaches smoothly to the disks. Although a clear division is seen in the
\pv\ diagram (Fig.~\ref{figure:HIPV}a) the two are not clearly
divided in moment maps.

\subsection{The Southern Tail}
\label{section:ResultsSouthernTail}
\indent

The southern tidal tail describes a gradual but pronounced arc about 90~kpc 
long which dominates the maps (Fig.~\ref{figure:HIMomentMaps}). It contains 
$\sim14$ times more \HI\ than the northern tail, even though the two appear
similar optically. It appears continuous, although some clumping is visible in
Fig.~\ref{figure:HIMomentMaps}a.

The optical width of the tail is quite narrow compared to \HI. Gaussian fits
to \HI\ profiles across the tail give ${\rm FWHM}\approx100\arcsecintxtE$.

The \HI\ flux peaks at $\sim1.59\,\JypbkmpsE$ ($\baseexpE{9.9}{20}\,\psqcmE$)
close to the tip of the tail. This is quite a pronounced concentration 
containing an \HI\ mass of $\sim\baseexpE{8.2}{8}\,\MsolarE$ or $\sim30\%$ 
of the total for the tail and was known as a distinct \HI\ 
region in the maps by van der Hulst (1979).

The southern \HI\ tail covers a continuous velocity range of 
$\sim1590-1784~\kmpsE$, with the velocity changing smoothly along the tail,
peaking at $\sim1773\,\kmpsE$. Towards the apparent concentration at the tip
of the tail, the velocity changes rapidly towards a minimum of 
$\sim1634\,\kmpsE$. The iso-velocity contours change orientation suggesting 
that the tail bends sharply toward the end, in agreement with a sharp bend
in the optical image. Hence this concentration might be just the result of
projection in a tail with varying orientation.

\subsection{The Northern Tail and Extension}
\label{section:ResultsNorthernTail}
\indent

Compared to the southern tail, the northern tail contains less \HI\ 
($\sim\baseexpE{1.6}{8}\,\MsolarE$), is shorter and covers a smaller 
radial velocity range (1550--1620\,\kmps). Its velocity decreases steadily
with distance from the disks.

There is a notable \HI\ gap between the northern tail and the disks
(see Fig.~\ref{figure:HIMomentMaps}b and Fig.~\ref{figure:HIChannelMaps}).
However, the \pv\ diagrams (Fig.~\ref{figure:HIPV}) suggest that a faint 
connection may exist. This gap contrasts with this tail's optical connection
to the disks and the continuous \HI\ in the southern tail.

Fig.~\ref{figure:HIMomentMaps}a shows \HI\ emission slightly west of the tip
of the northern tail, not corresponding to any optical feature and not noted
in any previously published \HI\ maps. It has an \HI\ mass of 
$\sim\baseexpE{1.8}{7}\,\MsolarE$ and a velocity range of 1550--1580\,\kmps.
Its closeness to the tip of the northern tail suggests that this is a westward
extension of a curved northern tail.

\subsection{The dwarf companion ESO~572--G045}
\label{section:ResultsESO572G045}
\indent

Our \HI\ maps (see Fig.~\ref{figure:HIMomentMaps}a,b) show a clear \HI\
detection of the small edge-on spiral galaxy ESO~572--G045 (Karachentsev
\etal\ 1993; see also Table~\ref{table:TheGroup}). Optically it has a size
of $\sim64\arcsecintxtE\times7\arcsecintxtE$. It is $\sim$16\farcm5
(91~kpc) from the centre of the Antennae, and $\sim$5\arcmin--6\arcmin\
south-west of the end of the southern tail.

In \HI\ the galaxy is slightly extended north-south, but unresolved
east-west, matching quite well the optical appearance. It has a clear and
smooth velocity gradient along the north-south major axis, ranging from
1650--1750\,\kmps\ (north to south), like an edge-on view of a rotating disk.
Its velocity also matches the adjacent portions of the Antennae southern tail.
There is no visible bridge of material joining this galaxy to the tidal tail,
and also no visible distortion of the tail in the direction of this galaxy.

The \HI\ mass of ESO~572--G045 is $\sim\baseexpE{1.4}{8}\,\MsolarE$, or $3~\%$
of that in the Antennae. Its systemic velocity is $\sim1700\,\kmpsE$.

\subsection{Radio Continuum Emission}
\label{section:ResultsRadioContinuum}
\indent

The ATCA 20\,\cm\ continuum map (Fig.~\ref{figure:Antennae20_13CM_Overlay}a)
corresponds to those published by Condon (1983; 1987), and by Hummel \& van der
Hulst (1986), all using VLA data. However, the 13\,\cm\ map obtained with the
ATCA (Fig.~\ref{figure:Antennae20_13CM_Overlay}b) is the first ever published.

The continuum sources seen in and around the Antennae are summarised in
Table~\ref{table:ContinuumSources}. The images show no emission outside of
the vicinity of the galactic disks, except for a number of point sources.
No continuum emission from the dwarf galaxies was detected.

Fig.~\ref{figure:Antennae20_13CM_Overlay} shows continuum concentrated in the
northern starburst ring and in the disk overlap region, with some additional
extended emission. The emission is well correlated with the location of star
formation activity as seen through optical, \Ha\, and infrared mapping
(Whitmore \etal\ 1999; Amram \etal\ 1992; Mirabel \etal\ 1998, respectively).
The continuum appears generally more extended than these regions.

The major peak in the radio continuum at both 20\,\cm\ and 13\,\cm\ occurs in
the disk overlap, with secondary peaks near the southern nucleus and in
the western portion of the starburst ring, as for the mid-infrared (Mirabel
\etal\ 1998). Sources~A, B, and C correspond to the three molecular gas
concentrations found in CO by Stanford \etal\ (1990).

Source~A, in the overlap region, is particularly bright. It corresponds to
`Region~C' of Whitmore \& Schweizer (1995), and to the `East Clump' of Stanford
\etal\ (1990). Higher-resolution 13\,\cm\ maps (not shown), with a fitted
Gaussian restoring beam rather than the larger circular one of
Fig.~\ref{figure:Antennae20_13CM_Overlay}, show this region resolved into three
or four smaller ones, apparently corresponding to the southern-most group
of emission regions seen in \Halpha\ by Amram \etal\ (1992).

Source~B is at the catalogued central or nuclear position of NGC~4038
(Table~\ref{table:TheGroup}), while Source~C is $\sim6\arcsecintxtE$ west of
the nucleus of NGC~4039.

The 20\,\cm\ map has slightly lower resolution than the 13\,\cm\ map (see 
Fig.~\ref{figure:Antennae20_13CM_Overlay}) and shows more extended emission,
seen largely as a number of faint `protrusions' from the Antennae disks, 
some of which are noise. Because of the higher resolution, the 13\,\cm\ map
resolves more emission peaks, particularly in the southern disk, and shows
two depressions which are not obvious at 20\,\cm.

Fig.~\ref{figure:Antennae20_13CM_Overlay} also shows some continuum emission 
coming from the base of the southern tail, which is rather bright optically
and in \HI\ (Fig.~\ref{figure:HIMomentMaps}b). There is no trace of radio
continuum emission elsewhere in either tail. 

We measure the total continuum emission from the disks as $\sim$498~mJy at
$1413\,\MHzE$ (20\,\cm), and $\sim$311~mJy at $2378\,\MHzE$ (13\,\cm). This
includes the emission at the base of the southern tail. The 20\,\cm\ figure
compares with the $472\pm23\,\mJyE$ at $1465\,\MHzE$ by Condon (1983),
$572\,\mJyE$ at $1490\,\MHzE$ by Condon (1987), and $486\pm20\,\mJyE$ at
$1465\,\MHzE$ by Hummel \& van der Hulst (1986). 

\section{Discussion}
\label{section:Discussion}

\subsection{\HI\ Mass Distribution}
\label{section:DiscussionHIMass}
\indent

The total \HI\ mass of the Antennae, as measured with the ATCA, is
$\sim\baseexpE{5}{9}\,\MsolarE$. About 46\% of the \HI\ mass is in the disks,
51\% in the southern tail and only 3\% in the northern tail (see
Table~\ref{table:HIstats}). Van der Hulst (1979) estimated the corresponding
components as 40\%, 52\% and 8\%. The column densities in the disks are much
higher than those in the tails. Our measurements agree with the general
trends found previously (\eg\ Hibbard \& van Gorkom 1996), whereby gas is
depleted in galactic disks as their merger progresses, leaving most of it in
the tidal tails. For the Antennae, at an early stage of merging, the tails
already contain about half of the \HI.

The southern tail is about twice as long optically as the northern tail, but
otherwise they are optically similar. In \HI\ this is quite different, with the
southern tail containing $\sim15$ times as much \HI\ as the northern tail. 
Additionally, the northern tail is not connected to the disks in \HI.

Wilson \etal\ (2000) mapped the molecular gas in the disks and find that about
half of the detected gas resides in five super-giant molecular cloud complexes
in the disk overlap. The remainder is in the galactic nuclei and the western
portion of the northern star-forming ring. However, the authors acknowledge,
based on single-dish observations by Gao \etal\ (1998), that they probably 
missed 60\% of the gas. More recently, the Gao \etal\ (2001) give a molecular
mass estimate of $\sim\baseexpE{1.38}{10}\,\MsolarE$ (adopting $D=19.2\,\MpcE$).
Combined with our \HI\ data, this implies a molecular-to-\HI\ mass ratio
of $\sim6:1$ for the disks alone and $\sim3:1$ for the whole Antennae.

Gao \etal\ (2001) made the first, tentative detection of CO at the tip of the
southern tail, with a molecular mass of $\sim\baseexpE{2}{8}\,\MsolarE$, and
hence a molecular-to-\HI\ ratio of $\sim1:4$.

Van der Hulst (1979) tried to overcome the large angular size of the Antennae,
and Westerbork's low sensitivity to east-west structure, by estimating \HI\
masses from multiple single-disk measurements by Huchtmeier \& Bohnenstengel
(1975). The derived total \HI\ mass is $\baseexpE{3.7}{9}\,\MsolarE$, much more
than the $\baseexpE{2.5}{9}\,\MsolarE$ which he measured by synthesis imaging,
but less than our measured  $\sim\baseexpE{5.1}{9}\,\MsolarE$ (including
ESO~572--G045), and the estimate of $\sim\baseexpE{5.3}{9}\,\MsolarE$ by
Huchtmeier \& Bohnenstengel. The component masses measured by van der Hulst
were $\baseexpE{1.0}{9}\,\MsolarE$ for the disks, $\baseexpE{1.3}{9}\,\MsolarE$
for the southern tail, and $\baseexpE{1.8}{8}\,\MsolarE$ for the northern tail
(adopting $D=19.2\,\MpcE$).

\subsection{The Galactic Disks}
\label{section:DiscussionDisks}

\subsubsection{Disk \HI}
\label{section:DiscussionDiskHI}
\indent

Both the \HI\ distribution and velocity pattern in the central area of the
Antennae (Fig.~\ref{figure:HIMomentMaps}~b,d) do not resemble the regular,
symmetric pattern seen in an isolated disk or spiral galaxy, and there is 
also no clear way to model it as the superposition of two disks. 

Our data suggests instead a) that disk \HI\ has been extensively kinematically
disturbed by disk collision or the preceding close encounter, b) that it is now
largely detached from the individual galaxies, with most of it no longer
following closely their original disk structure, and c) that it resides 
largely in a single structure between and surrounding the disks. Further, 
this structure is connected smoothly to the \HI-rich southern tidal arm.

Mihos \& Hernquist (1994) ran computer simulations of a variety of merging disk
galaxy pairs, including both those with and without large bulges. In the
simulations, a clear pattern was seen. Galaxies with no bulge experienced
nuclear gas inflows early in the merger, producing a steady moderate starburst
and depleting the gas prior to the final stages of the merger. However, the
presence of a substantial bulge inhibited this inflow, with a low initial
star-formation rate and minor gas depletion. The result for these systems was
coalescence of the two galaxies' still abundant gas late in the merger,
resulting in a massive burst of star formation and an ultraluminous infrared
merger.

This study sheds light on the possible origin of the observed \HI\ pool in the
Antennae, and their future evolution. The star formation consequences of this
are described further in Section~\ref{section:DiscussionStarFormation}.

Although there is an apparent pooling of the \HI, the dual-peaked spectra,
seen particularly over the central southern disk, indicate that in this region
at least there are two clouds overlapping along the line-of-sight, with
some of the \HI\ still attached to the southern disk.

\subsubsection{Disk Radio Continuum}
\label{section:DiscussionDiskContinuum}
\indent

Comparison of our ATCA 20 and 13 cm radio continuum data 
(Fig.~\ref{figure:Antennae20_13CM_Overlay}) with VLA maps at 20 and 6 cm
(Hummel \& van der Hulst 1986) does not reveal many new features.

The 13 cm continuum map (beam = 9\farcs5) resembles a smoothed version
of the high-resolution 6\,\cm\ and 20\,\cm\ VLA maps (beam = 6\arcsec). The
latter resolve greater detail in the disks, \eg\ more emission peaks in the
eastern half. The faint emission at the base of the southern tail is not seen
in these maps, but is clearly detected in the low-resolution VLA 20-cm map.

The low-resolution 20\,\cm\ map by Hummel \& van der Hulst (1986) has the same
sensitivity and resolution as our Fig.~\ref{figure:Antennae20_13CM_Overlay}a,
allowing a direct comparison. The measured fluxes agree well within the errors.
Our image includes slightly more extended emission at faint levels, but most 
of it is likely to be noise.

The well-resolved emission peaks in our radio continuum maps, particularly 
in the 13\,\cm\ image (Fig.~\ref{figure:Antennae20_13CM_Overlay}b), agree 
with those in the \Ha\ map by Amram \etal\ (1992). Our resolution is too low
to allow a detailed comparison of the radio continuum and \Ha\ peaks.

With radio continuum closely matching optical \HII\ regions and \Ha\ maps, the
Antennae agree with the correlation between star formation and the \Ha, FIR,
and radio, noted for example by Sanders \& Mirabel (1985).

\subsection{The Tidal Tails}
\label{section:DiscussionTidalTails}

\subsubsection{Modelling The Tails}
\label{section:DiscussionTailModelling}
\indent

Toomre \& Toomre (1972) performed pioneering computer simulations which clearly
demonstrated for the first time that galactic tails originate through tidal
interaction. However, these simulations entirely neglected the self-gravity of
the disk material as well as dispersion effects, which have been considered
in more recent simulations. The omission of simulated velocity fields in many
theoretical papers on galaxy interactions often renders impossible a direct
kinematic comparison between \HI\ observations and model predictions.

Fortunately, van der Hulst (1979) obtained and published graphically the
velocity information not published by Toomre \& Toomre (1972), and used it to
compare observed \pv\ diagrams to predicted ones. After rescaling of dimensions
to fit the data, the match was quite close.

Computer simulations (\eg\ Appendix VIII of Hibbard 1995; Hibbard \& Mihos
1995) predict that material in a tail will be ejected along many different
trajectories, with a large-scale filament-like structure due to the similarity
of trajectories of neighbouring material, rather than being a stream of
material along a common trajectory. In this way, it resembles a density wave,
with the wave initially moving outward from the galactic disks, and in most
cases later falling back inward under gravity. Consequently, the material
velocities tend to be at a large angle to the orientation of the tail, and 
disk galaxy models do not apply to tidal tails.  

More recently, Barnes (1988) carried out tree-code N-body simulations of
interacting disk galaxies, incorporating full self-gravity and a
bulge/disk/halo model. The first model made was a look-alike of the Antennae,
for which a number of figures were published, including a \pv\ diagram along
the orbital plane, approximately north-south (see his Fig.~3d). This is roughly
a modelled equivalent of our Fig.~\ref{figure:HIPV}b, but showing all material
originating from the disk rather than just \HI. These two figures bear a very
close resemblance, except that the observed southern \HI\ tail is much more
prominent than the northern one, whereas the model diagram shows them fairly
comparable in size.

Hibbard \& Mihos (1995) performed similar simulations based on earlier VLA \HI\
data for the merging pair NGC~7252, which closely resembles the Antennae. The
observed \HI\ \pv\ diagram (their Fig.~1) closely resembles our observed ones
for the Antennae (Fig.~\ref{figure:HIPV}). That work particularly mentioned
the final destination of material in tidal tails (their Figs.~5, 7). 
The bulk of tail material has an elliptical trajectory taking it back towards
the disks, much of it quite rapidly. Some, however, may gain escape velocity,
with material further from the disks also having higher velocities.

By analogy, it can be inferred that the outer-most material in the southern
tail of the Antennae (including the concentration at the tip), may take quite
some time to fall back, if it does at all. However, the turn-around in radial
velocities in the southern tail indicates that it is substantially bent along
the line-of-sight `towards' the disks, suggesting that the tip is falling 
back. This interpretation, though, is unlikely to be correct since computer
simulations show that the energy of tail material increases with distance 
from the center, and thus outer portions of the tail will take longer to fall
back (J. Hibbard, private communication).

\subsubsection{Optical versus \HI\ Emission in the Tidal Tails}
\label{section:DiscussionTailHIOpticalDifferences}
\indent

There is no obvious spatial displacement between the optical and \HI\ emission
in the tails (Fig.~\ref{figure:HIMomentMaps}) as has been observed in other
interacting systems, \eg\ Arp~157, Arp~220, and Arp~299 (see Hibbard, Vacca and
Yun 2000). The primary differences are: {\bf(1)} the widths of the tails,
{\bf(2)} the highly visible \HI\ pool at the end of the southern tail, {\bf(3)}
the greater length of the southern tail in \HI, and {\bf(4)} the absence of
\HI\ at the base of the northern tail. The latter might be due to ionisation by
emission from the disks as suggested by Hibbard, Vacca and Yun (2000).

The \HI\ in the tails is clearly more extended laterally than the stellar 
emission. Short-exposure optical images (\eg\ Fig.~1) show both tidal tails
to be fairly narrow with a width of 30\arcsec--40\arcsec\ (2.8--3.7~kpc),
much less than the \HI\ width of $\sim$100\arcsec\ (9.3~kpc). Deeper optical
images of the Antennae (Schweizer 1978), however, show both tails more being
extended.

The strong stellar emission from the Antennae tails is somewhat unusual since
most tidal tails are optically faint. This general trend may come about because
material which is (pre-interaction) further from the galactic centre, is less
tightly bound gravitationally, and more affected by intergalactic tidal
forces, and thus generally contributes most of the material in tidal tails
(\eg\ Toomre \& Toomre 1972). This and the fact that in most galaxies the gas
is much more extended than the stellar component, leads to tails initially
containing a lot of gas and relatively few stars. Furthermore, the rapid
decrease in the gas density of tidal arms, due to differential expansion after
their creation, should result in a corresponding decrease in star-formation
rates, and hence a drop in optical emission (Mihos, Richstone \& Bothun 1991).
The fact that the Antennae do not show low optical brightness, suggests that
the material in the tails derives from a more inner part of the disks than what
is often seen in other galaxies.

\subsubsection{No Continuum Emission in the Tidal Tails}
\label{section:DiscussionTailContinuum}
\indent

Although the disks show continuum emission, we have found none from either of
the tidal tails, except at the base of the southern tail. Since radio continuum
emission is generally produced extensively in star-forming regions, \eg\ by 
hot gas or shocks in supernova remnants, its absence in the tails implies a low
level of star formation.

The absence of detectable radio continuum in tidal tails is also quite common
for other observed interacting systems, and agrees with predictions by Mihos,
Richstone \& Bothun (1991) and others, who have undertaken computer simulations
of interaction-induced starbursts, modelling galactic close encounters and the
resulting collisions of giant molecular clouds and inwards gas flows, believed
to be the means by which starbursts are produced in interacting galaxies.
Mihos, Richstone \& Bothun (1991) studied star-formation rates for
pairs of disk galaxies undergoing encounters (but not mergers) at a number of
separations and relative orientations. They found that in some cases
close encounters will actually reduce star formation, by throwing clouds out of
the disk and reducing their density, without producing many collisions. 
However, prograde encounters (both disks rotating in the same sense as their
orbit about one another, \eg\ the Antennae), tend to produce quite strong
starbursts and extensive transfer of material between the disks. Such
encounters could also produce starbursts along the leading edges of tidal
tails. However, such tails expand rapidly and the star formation rate drops
considerably after a couple $\times10^8{\rm~years}$ (Mihos \etal\ 1991), so
that tails will generally have no detectable star formation activity when
observed well into a close encounter or merger.

\subsection{Southern-Tail Tidal Dwarf Galaxy}
\label{section:DiscussionTailDwarf}
\indent

The enhanced \HI\ column density at the tip of the southern tail coincides with
the location of a young ($\sim\baseexpE{3}{8}$ years) star-forming region,
suspected to be a dwarf galaxy formed during the interaction (Schweizer 1978;
Mirabel, Dottori \& Lutz 1992). The enhancement, by a factor of $\sim4.5:1$
compared to the tail's minimum can at least in part be explained as a 
projection effect due to the bending of the tail and hence a deeper line 
of sight. However, such a large ratio requires a substantial bending angle,
hence there is likely to be an enhanced local gas density at the location 
of the star-forming region: a point already made by van der Hulst (1979).

Sizeable concentrations in tidal arms, many of which appear self-gravitating,
have been found in \HI\ studies of a number of interacting galaxies, with
typical \HI\ masses of $\sim\baseexpE{5}{8}-\baseexpE{6}{9}\,\MsolarE$, the
resulting overall population of bound dwarf galaxies being perhaps half of all
dwarfs (Sanders \& Mirabel 1996).

The formation of dwarf galaxies is predicted by computer simulations of galaxy
interaction, such as those by Barnes \& Hernquist (1992), and the more complete
model, including gas dynamics, by Barnes \& Hernquist (1996). In models,
self-gravitation of some material ejected in tidal arms, may be enough to
overcome expansion of the arm over time, forming bound objects. These tend to
be gas-rich since their gas and stars have slightly different origins, with the
gas being more compact and so surviving tidal disruption. Frequently a number
of bound objects are produced, generally in the outer parts of the tail, due to
the lower mass needed to avoid disruption there.

Elmegreen \etal\ (1993) used computer models of IC~2163/NGC~2207 to show that
extended disks of gas can lead to a massive concentration of gas at
the end of a tidal arm during interaction. They also showed that if the mass of
the second galaxy is similar to, or larger than, the host galaxy, then
eventually massive clouds can be ejected from the host galaxy, leading to
self-gravitating dwarf galaxies like those observed by the same authors using
VLA \HI\ observations. According to Elmegreen \etal, the lack of an extended
gas distribution in the models of Barnes \& Hernquist (1992), prevented the gas
pool forming in their simulations. With the Antennae galaxies merging, and of
apparently similar masses, they should be sufficient to produce a pool along a
tail, or attached or detached dwarf galaxies.

\subsection{The dwarf companion ESO~572--G045}
\label{section:DiscussionESO572G045}
\indent

The galaxy ESO~572--G045 is much smaller and less massive than the Antennae
and is classified as a spiral dwarf galaxy. It has an \HI\ radial velocity
$\sim1688\,\kmpsE$, similar to that of the southern tidal tail, and a small
apparent distance from it.

The association between ESO~572--G045 and the Antennae could be found by
searching it for signs of recent interaction (\eg\ starburst activity). If this
has not occurred then almost certainly the dwarf is not involved in the
Antennae merger. If interaction has occurred, then either the dwarf was a
pre-existing galaxy now tidally disturbed, or it is a bound object produced
from tidal debris from the merger (\eg\ Elmegreen \etal\ 1993; Hernquist 1992,
1996). Given its apparent spiral structure, creation from debris appears a less
likely explanation. With an \HI\ mass alone of
$\sim\baseexpE{1.4}{8}\,\MsolarE$, its mass would be at the high end for bound
objects (Barnes \& Hernquist 1992), but still reasonable. Unfortunately, to
date there is no available information which would resolve this question.

The small size of the object relative to the beam allows only an approximate
dynamical mass estimate from \HI\ velocities. It is edge-on, hence its
maximum rotational velocity is half of the \HI\ range ($\sim50\,\kmpsE$).
The corresponding radius appears to be $\sim32\arcsecintxtE$ (3\,\kpc). 
The following equation (Giovanelli \& Haynes 1988) estimates the total 
enclosed mass, assuming a spherical distribution:

\begin{equation}
\label{equation:MvsRV_spherical}
M_T(r) = \frac{rV^2(r)} {\kpcE~(\kmpsE)^2} \times \baseexpE{2.33}{5}\,\MsolarE~~~~,
\end{equation}
from which we derive a dynamical mass of $\sim\baseexpE{1.74}{9}\,\MsolarE$.
These estimates imply that \HI\ makes up about 8\% of the dwarf's mass, roughly
in the range observed in larger spiral galaxies, and supporting the suggestion
of spiral structure.

The inner-most distance from the center of a globular cluster or orbiting
satellite, at which tidal stripping occurs, is the tidal radius ($r_t$),
which we take as the Jacobi limit:

\begin{equation}
\label{equation:TidalRadius}
r_t = R {\left(\frac{M}{3M_g}\right)}^{1/3}~~~,
\end{equation}
where $M$ is the total mass of the satellite, $M_g$ that of the parent galaxy 
and $R$ their separation (see \eg\ Pisano \& Wilcots 2000).
Taking the velocity minima and maxima as the extreme velocities in the Antennae
disks, and the \HI\ peak as the center of mass, leads to projected rotation
velocity of $\sim228\,\kmpsE$ at a radius of $\sim12.5\,\kpcE$, and a
very rough Antennae disk mass estimate of $\sim\baseexpE{1.5}{11}\,\MsolarE$.
Treating ESO~572--G045 as a satellite of the Antennae disks ($R \ge91$~kpc)
gives a tidal radius of $r_t\ge14$~ kpc, much larger than its measured \HI\
radius. So it is unlikely that ESO~572--G045 is prone to disruption.

\subsection{Star Formation Patterns}
\label{section:DiscussionStarFormation}
\indent

The Antennae starburst is optically spectacular and has a total infrared
luminosity of $\sim10^{11}~\LsolarE$ indicating major starburst activity.
However, Gao \etal\ (2001) concluded that the starburst is modest considering
the high molecular gas content of the system ($\baseexpE{1.38}{10}\,\MsolarE$).

Computer models of the Antennae have shown that it may result from the 
prograde encounter of two disk galaxies. This has 
been found (\eg\ for the non-merging case, by Mihos \etal\ 1991) to generally
produce the strongest starburst. In the same study, closer encounters also
strengthen the starburst at least for non-mergers. The Antennae encounter and
the starburst within it thus agrees with the available models.

Mihos \& Hernquist (1994) modelled star formation patterns in merging pairs 
of disk galaxies, concluding that galaxies with large bulges would delay their
major starburst activity until the final stages of merger. At this point a
combined central gas pool would produce vigorous starbursts (and a high 
infrared luminosity). From our \HI\ studies it appears that such a pool of
cool gas is forming in the Antennae.

These observations suggest that the Antennae are an example of the delayed
action for bulge galaxies modelled by Mihos \& Hernquist, with the current
starburst being an early stage, and that the scene is being set for a fairly
spectacular final merger of the disks at a later date.

\section{Conclusion}
\label{section:Conclusion}
\indent

Our \HI\ images of the Antennae have higher sensitivity and resolution than
those by van der Hulst (1979), revealing more \HI\ mass and structure as well 
as some new features not previously observed. 

Our radio continuum images include the first published 13\,\cm\ image, and a
new 20\,\cm\ image corresponding to previously published ones by Condon (1983,
1987) and Hummel \& van der Hulst (1986). No continuum polarisation has been 
detected.  The radio continuum maps closely resemble the \Ha\ maps by Amram 
\etal\ (1992). The distribution approximates the optical star-forming regions,
\eg\ seen in HST images. Sources in the overlap region match the locations of
giant molecular clouds seen in CO by Stanford \etal\ (1990). These similarities
suggest that the continuum results largely from star formation. The lack of
continuum emission in the tidal tails indicates that recent and on-going star
formation in the Antennae is basically confined to the disks.

The presence of a central \HI\ gas pool along with a modest disk starburst (Gao
\etal\ 2001) and computer simulations (Mihos \& Hernquist 1994) suggest that
the current starburst may be a weak precursor to a far larger one occuring much
later in the merger.

The southern tidal tail contains abundant \HI\ mass 
(\baseexp{2.5}{9}\,\Msolar), similar to the galactic disks
(\baseexp{2.3}{9}\,\Msolar), but less concentrated. This tail is connected
smoothly to the disks, although \pv\ diagrams show a clear breakaway point at
1590\,\kmps. Definite \HI\ clumps exist along the tail's length. It has a
substantial \HI\ width of up to $\sim100\arcsecE$ ($\sim9\,\kpcE$), larger than
the optical width. There is no spatial displacement between optical and \HI\
emission.
The tip of the southern tail contains a large \HI\ concentration containing
$\sim\baseexpE{8.2}{8}\,\MsolarE$ or $\sim30\%$ of the tail's \HI. At this point
the tail also bends sharply both in terms of velocity and shape, and the \HI\
column density is higher by a factor of $\sim4$. This coincides in position
with the young tidal dwarf galaxy, observed optically by Mirabel \etal\ (1992).
Dwarf galaxies have previously been produced as a tidal feature by computer
simulations, along with pools of material at the ends of tidal arms (Barnes \&
Hernquist 1992, 1996; Elmegreen 1993), and they have also been observed in 
other interacting galaxies (\eg\ Duc \& Mirabel 1994, 1998). Although it is not
known that any of the \HI\ pool in the Antennae is associated with the dwarf,
it appears likely they are associated, as seen in the simulations.

The smaller northern tail contains only \baseexp{1.8}{8}\,\Msolar\ of \HI\
and has no detectable \HI\ connection to the disks, which may be partly due
to ionisation effects suggested by Hibbard, Vacca \& Yun (2000). This tail's
\HI\ extends much further lengthwise than its starlight, unlike the southern
tail, and it curves toward the west along a faint extension newly detected
in our \HI\ data and not visible optically.

The central \HI, while it follows broadly the outline of the optical disks, 
has a detailed velocity structure and spatial distribution which bear no close
resemblance to those normally observed in disks. The \HI\ appears to have been
so kinematically disturbed that it is largely decoupled from the stellar
component. The overall velocity range is a high $\sim640\,\kmpsE$ and the \HI\
is substantially pooled in the region between the disks, peaking midway between
them, unlike optical images. While generally there is only one velocity
component along the line-of-sight, part of the disk \HI, close to the southern
nucleus, contains two components and thus a high dispersion, up to 150\,\kmps.

Comparison of \HI\ mass measurements to previous CO measurements by Wilson
\etal\ (2000) and Gao \etal\ (2001) gives an estimated molecular-to-atomic
gas ratio of $\sim3:1$ for the whole system. The difference between the
physical sizes and \HI\ contents of the tails, suggests a difference in
structure or gas content between the two progenitor galaxies, although
asymmetry in the encounter geometry may be responsible.

Our \HI\ images also show the nearby dwarf galaxy ESO~572--G045, which is
south-west of the end of the southern tail, and $\sim$91~kpc from the Antennae
centre. The \HI\ matches the optical appearance of the galaxy, an edge-on disk
$\sim$1\arcmin\ in width. Using the \HI\ velocity field we estimate a systemic
velocity of $\sim1700\,\kmpsE$ and a rotational velocity of $\sim50\,\kmpsE$,
resulting in a dynamic mass of $\sim\baseexpE{1.74}{9}\,\MsolarE$, compared to
an \HI\ mass of $\sim\baseexpE{1.4}{8}\,\MsolarE$. This gives an 8\% \HI\ 
content which appears reasonable. The \HI\ mass of ESO~572--G045 is 3\% that 
of the Antennae. The dwarf's closeness, and quite similar velocity, to the
adjoining southern tail of the Antennae, indicates that it is a companion
galaxy. It appears likely to have been involved in the interaction between
the Antennae galaxies, but no major disturbances are observed. Its tidal 
radius is $\ge14\,\kpcE$, so it is unlikely to be prone to disruption.

\section{Acknowledgements}
\label{section:Acknowledgements}

\begin{itemize}

\item We would like to thank both the anonymous referee, and John Hibbard, for
their many valuable suggestions, which have been incorporated into this paper.

\item Some figures presented use data from the National Aeronautics and Space
Administration (NASA)'s {\it SkyView} facility
(\URL{http://skyview.gsfc.nasa.gov}) located at NASA Goddard Space Flight
Center. In particular, use has been made of Digitised Sky Survey (DSS) images
based on photographic data obtained using the UK Schmidt Telescope.

\item Various information used in preparing this paper, was taken from the
NASA/IPAC Extragalactic Database (NED), operated by the Jet Propulsion
Laboratory, California Institute of Technology, under contract with NASA.

\end{itemize}



\begin{table*} 
\caption[Properties of NGC~4038/9 and ESO~572--G045.]{
\label{table:TheGroup} 
         Properties of NGC~4038/9 (`The Antennae') and ESO~572--G045.  }
\begin{tabular}{lcccccl}
\hline
Name          & \multicolumn{2}{c}{J2000 Position}                     
              & Optical Velocity         
	      & ${\rm B}^o_T$\notenum{*} 
	      & Galaxy Type \\
              & RA                         
	      & DEC                       
	      & (heliocentric) \\
\hline
NGC~4038      & \hms{12}{01}{52.8}\notenum{1} 
              & \dms{-18}{52}{05}\notenum{1} 
	      & $1663\pm20~\kmpsE~^{(2)}$  
	      & 10.59\notenum{2}   
	      & SB(s)m(pec)\notenum{2} \\
NGC~4039      & \hms{12}{01}{53.6}\notenum{1} 
              & \dms{-18}{53}{12}\notenum{1} 
	      & $1655\pm28~\kmpsE~^{(2)}$  
	      & 10.69\notenum{2}   
	      & SA(s)m(pec)\notenum{2} \\
ESO~572--G045 & \hms{12}{01}{09.4}\notenum{3} 
              & \dms{-19}{04}{49}\notenum{3} 
	      & unknown                    
	      &                    
	      & Scd\notenum{3} \\ 
\hline
\multicolumn{6}{l}{
Linear distance scale ($D=19.2~\MpcE$): $1\arcminintxtE \approx 5.6~\kpcE$.
\vspace{0.5cm}} \\
\multicolumn{6}{l}{\notenum{*} The photoelectric total magnitude, corrected 
  for galactic and internal absorption, and redshift. Refer to \notenum{2}}. \\
\multicolumn{6}{l}{\notenum{1} Stanford \etal\ (1990)}. \\
\multicolumn{6}{l}{\notenum{2} de Vaucouleurs \etal\ (1991).} \\
\multicolumn{6}{l}{\notenum{3} Karachentsev, Karachentseva \& Parnovskij 
 (1993). }
\end{tabular}
\end{table*}

\begin{table*} 
\caption[Details of Antennae Observations]{
\label{table:ObsDetails}
         Details of the radio observations.}
\begin{tabular}{ll}
\hline
Instrument               & Australia Telescope Compact Array (ATCA) \\
Array configurations     & 6C (1996~July~24);~~~1.5A (1996~Oct~25);~~~750A (1996~Nov~12; 13);~~~375 (1998~Mar~28) \\
Baseline coverage        & 30.6--6000~\m, 75 baselines (56 unique) \\
Pointing centre          & RA=\hms{12}{01}{40}, DEC=\dms{-18}{55}{00} \\
Frequency configuration  & IF1 (20\,\cm): centre 1413~\MHz, bandwidth 8~\MHz, 512 channels, 2 polarisations \\
                         & IF2 (13\,\cm): centre 2378~\MHz\ (6C, 1.5A, 750A arrays) or 2496~\MHz\ (375 array) \\
                         & ~~~~~~~~~~~~~~~~~~bandwidth 128~MHz, 32 channels, 4 polarisations \\
Primary beam FWHM        & $33\arcminintxtE$ (IF1 or 20\,\cm), $21\arcminintxtE$ (IF2 or 13\,\cm) \\ 
Observing time per run   & 11--12 hours overall; $\sim8.5-9.5$ hours on source [3.8/2.3 hours for second 750A run] \\
Total observing time     & 50 hours overall; 39 hours on source \\ 
Flux calibration         & 10~min per run, using PKS~1934--638 (14.88~\Jy\ in IF1, 11.5~\Jy\ in IF2) \\
Phase calibration        & alternately 30~min on-source, 3~min on PKS~1151--348 (6.1/4.1~\Jy) or 1127--145 (5.0/4.65~\Jy) \\
Calibrated Bandwidth     & IF1 (20\,\cm): 6.8~\MHz\ (1436~\kmps\ as \HI) ; IF2 (13\,\cm): 100~\MHz\ \\
Image Weight; r.m.s. noise; Beam & \HI\ : `robust=1'; 1.0~\mJypb\ vs 0.95~\mJypb\ theoretical; $42\arcsecintxtE$ beam \\
                         & 20\,\cm\  : `uniform'; 0.23~\mJypb\ vs 0.25~\mJypb\ theoretical; $16\arcsecintxtE$ beam \\
                         & 13\,\cm\  : `uniform'; 0.056~\mJypb\ vs 0.039~\mJypb\ theoretical; 9\farcs5 beam \\
\hline
\end{tabular}
\end{table*}

\begin{table*} 
\caption[\HI\ Statistics for the Antennae]{
\label{table:HIstats}
\HI\ flux statistics for various regions of the Antennae group, following
primary-beam correction. The \HI\ velocity ranges are estimated from the
channel maps (see Fig.~\ref{figure:HIChannelMaps}). Mass percentages are
relative to the combined mass of NGC~4038/9 and ESO~572-G045.
}
\begin{tabular}{lcccccccc}
\hline
Region                    & \HI\ Velocities & \HI\ Flux & \HI\ Mass         & Fraction    & Peak Integrated  & \multicolumn{2}{c}{J2000 Peak Position } \\
                             & (\kmps)      & (\Jykmps) & (\Msolar)         & of total    & Intensity (\Jypbkmps & RA ($\pm2.5\arcsecintxtE$) & DEC ($\pm2.5\arcsecintxtE$)\\
                          &                 &           &                   &             & and $\exponE{20}\,\pcmcmE$) \\
\hline
Galactic disks               & 1340--1945   & 26.3~~    & \baseexp{2.28}{9} & 45~\%~~     & 5.62 ; 35.0       & \hms{12}{01}{54} & \dms{-18}{52}{40} \\
Southern tail                & 1590--1784   & 28.8~~    & \baseexp{2.50}{9} & 49~\%~~     & 1.59 ; ~~9.9      & \hms{12}{01}{25} & \dms{-19}{00}{44} \\
~~-- Tip only                & 1634--1740   & 10.5~~    & \baseexp{8.23}{8} & 18.0~\%     & 1.59 ; ~~9.9      & \hms{12}{01}{25} & \dms{-19}{00}{44} \\
\multicolumn{2}{l}{Northern tail}           & ~2.05     & \baseexp{1.78}{8} & ~~3.5~\%    & 0.37 ; ~~2.3      & \hms{12}{02}{09} & \dms{-18}{49}{13} \\
~~-- Main section            & 1550--1622   & ~1.84     & \baseexp{1.60}{8} & ~~3.1~\%    & 0.37 ; ~~2.3      & \hms{12}{02}{09} & \dms{-18}{49}{13} \\
~~-- Extension               & 1550--1580   & ~0.21     & \baseexp{1.81}{7} & ~~~0.25~\%  & ~~0.15 ; ~~0.93   & \hms{12}{01}{57} & \dms{-18}{46}{08} \\
ESO~572--G045                 & 1640--1750   & ~1.61     & \baseexp{1.40}{8} & ~~2.7~\%    & 1.10 ; ~~6.9      & \hms{12}{01}{09} & \dms{-19}{04}{41} \\
\hline
\multicolumn{2}{l}{Combined totals:} \\
\multicolumn{2}{l}{~~Both tails}            & 30.9~~    & \baseexp{2.67}{9} & 53~\%~~     & 1.59 ; ~~9.9      & \hms{12}{01}{25} & \dms{-19}{00}{44} \\ 
\multicolumn{2}{l}{~~Southern tail + disks} & 54.9~~    & \baseexp{4.77}{9} & 94~\%~~     & 5.62 ; 35.0       & \hms{12}{01}{54} & \dms{-18}{52}{40} \\
\multicolumn{2}{l}{~~Both tails + disks}    & 56.9~~    & \baseexp{4.95}{9} & 97~\%~~     & 5.62 ; 35.0       & \hms{12}{01}{54} & \dms{-18}{52}{40} \\
\multicolumn{2}{l}{~~The whole system}      & 58.6~~    & \baseexp{5.08}{9} & 100~\%~~~   & 5.62 ; 35.0       & \hms{12}{01}{54} & \dms{-18}{52}{40} \\
\hline
\end{tabular}
\end{table*}

\begin{table*}
\caption[Radio Continuum Sources]{
\label{table:ContinuumSources}
A list of radio continuum sources (including previous identifications);
sources~A, B, C, and D are in the Antennae. The diffuse continuum emission
from the galactic disks is discussed elsewhere.  }
\begin{tabular}{lccccl}
Source Name    & \multicolumn{2}{c}{Peak Position } & \multicolumn{2}{c}{Peak Flux Density (\mJypb)} & Comments \\
/ Identifier   & RA                  & DEC               & 20\,\cm\        & 13\,\cm\   & \\
\hline
A              & \hms{12}{01}{55.0}  & \dms{-18}{52}{49} & 63.6~~~         & 22~~~~     & Overlap of disks (4 subregions) \\
B              & \hms{12}{01}{53.0}  & \dms{-18}{52}{06} & 32~~~~~         & 10.5~~     & NGC~4038 nucleus \\
C\notenum{1}   & \hms{12}{01}{53.3}  & \dms{-18}{53}{10} & 19\notenum{1}~~ & 6.5~     & 5\arcsec\ south-east of NGC~4039 nucleus \\
D              & \hms{12}{01}{48}~~  & \dms{-18}{55}{15} & 1.6~~           & 1.6~   & Midway between dwarf galaxies \\
E              & \hms{12}{01}{45}~~  & \dms{-18}{53}{05} & 3.4~~           & 3.4~   & West of the galactic disks \\
F              & \hms{12}{01}{09.7}  & \dms{-18}{43}{39} & 11~~~~~         & $<2$~~~~~~ & Extended ($\sim100\arcsecintxtE\times30\arcsecintxtE$) source (N--W)\\
TXS~1159-187   & \hms{12}{02}{12.3}  & \dms{-19}{00}{30} & 108~~~~~~~      & 43~~~~~   & Catalogued radio source\notenum{2} \\
TXS~1158-187   & \hms{12}{01}{18.3}  & \dms{-19}{02}{19} & 78~~~~~~        & 36~~~~~   & Catalogued radio source\notenum{2} \\
MS~1200.2-1829 & \hms{12}{02}{52.6}  & \dms{-18}{44}{57} & 5~~~~           &           & Known X-ray \& optical source\notenum{3} \\

\hline
\multicolumn{6}{l}{\notenum{1}~~This peak is poorly resolved due to
  contamination by neighbouring peaks, hence the value may not be accurate.}\\
\multicolumn{6}{l}{\notenum{2}~~Detected at 365~MHz by Douglas \etal\ (1996).}\\
\multicolumn{6}{l}{\notenum{3}~~$z=1.83$, $\sim$20~mag, see Gioia \etal\
  (1990), Maccacaro \etal (1991), Stocke \etal (1991). }

\end{tabular}
\end{table*}


\clearpage


\begin{figure*} 
\begin{tabular}{cc}
\mbox{\psfig{file=figure2e.ps,width=8.4cm}} &
\mbox{\psfig{file=figure2f.ps,width=8.4cm}}
\\
{\bf(e)} \HI\ moment-2 (velocity dispersion) &
{\bf(f)} \HI\ moment-2 (a closeup of {\bf(e)}) \\
\end{tabular}
\begin{center}
{\bf Figure~\ref{figure:HIMomentMaps} continued. }
\end{center}
\end{figure*}


\begin{figure*} 
\begin{tabular}{cc}
\mbox{\psfig{file=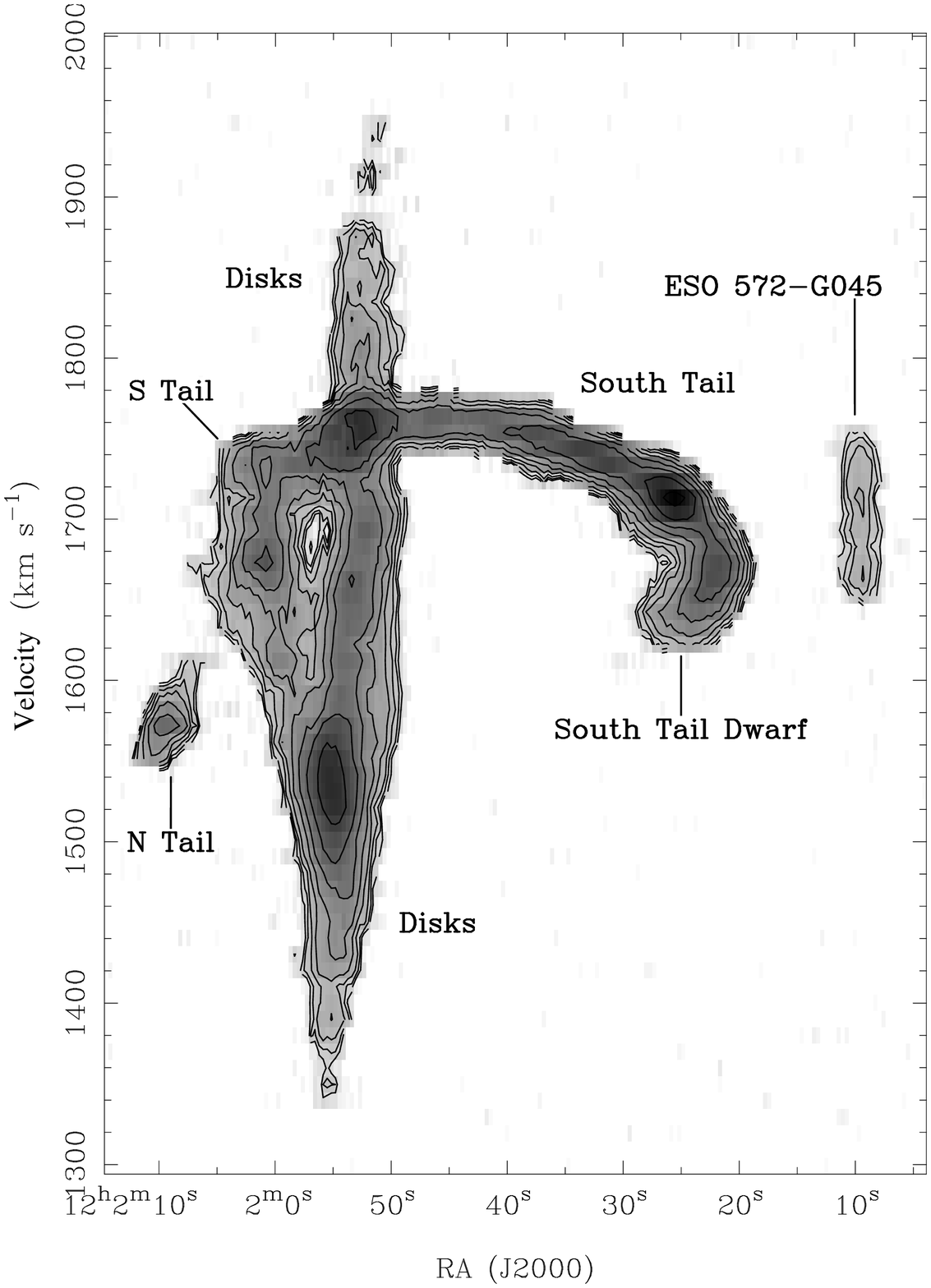,height=9.8cm}}&
\mbox{\psfig{file=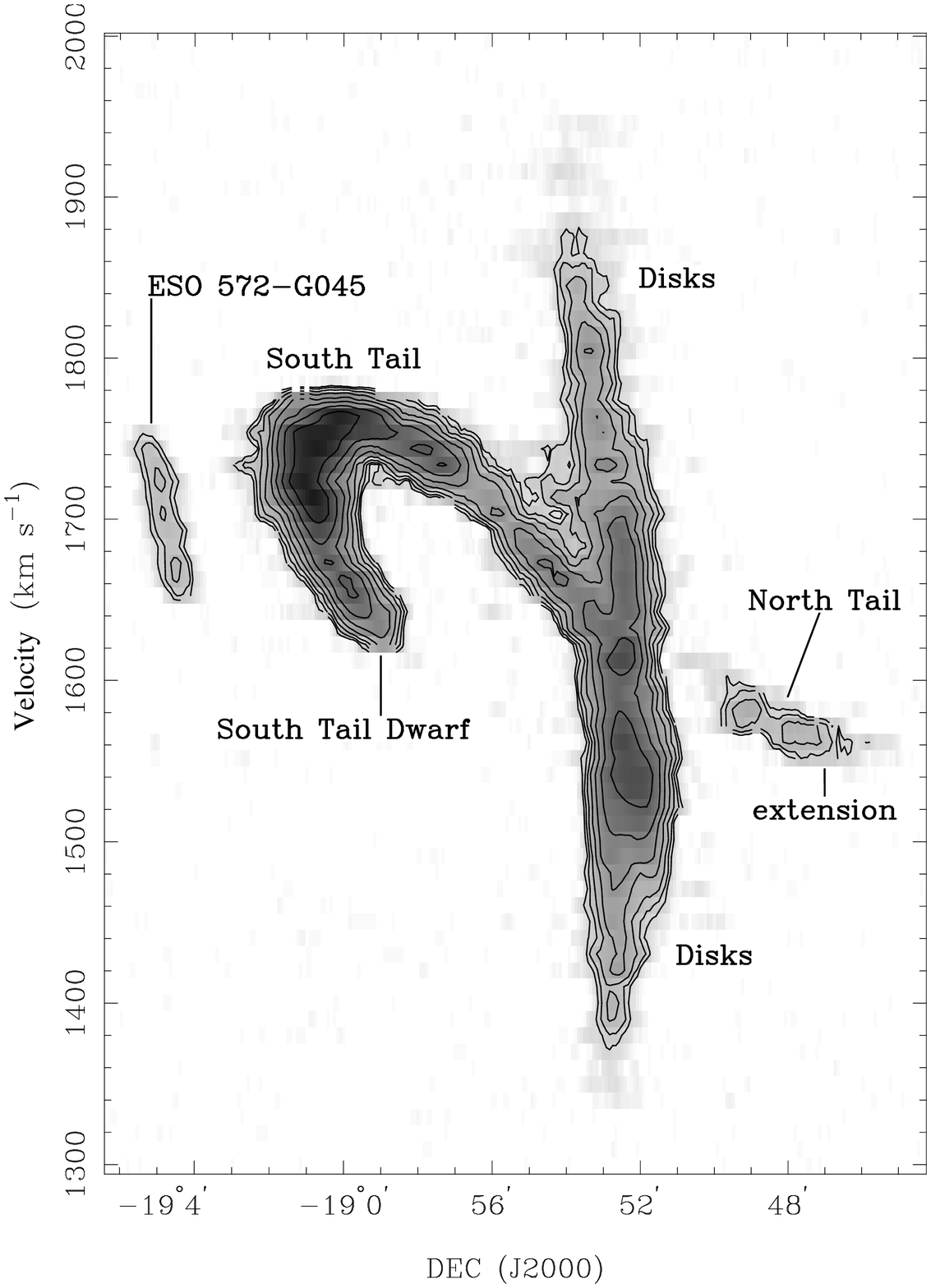,height=9.8cm}}\\
{\bf(a)} Projection onto (RA, velocity) axes &
{\bf(b)} Projection onto (DEC, velocity) axes
\end{tabular}
\caption[\HI\ \pv\ Diagrams for the Antennae]{
\label{figure:HIPV}
\HI\ position-velocity (\pv) diagrams, obtained by integrating the channel
maps along a spatial axis, after clipping to exclude noise: ~~{\bf(a)}~
integrated along the DEC axis, ~~{\bf(b)}~ integrated along the RA axis.
Contours (both cases): 0.2, 0.282, 0.4, 0.565, 0.8, 1.12, 1.6, 2.3, 3.2, 4.5,
$6.4\,\JypbE\,\arcsectxtE$, \ie\ powers of $\sqrt{2}$. Conversion:
$1\,\JypbE\,\arcsectxtE\approx\baseexpE{4.62}{5}~\MsolarE\invE{\left(\kmpsE\,\kpcE \right)}$. }
\end{figure*}


\begin{figure*} 
\begin{tabular}{cc}
\mbox{\psfig{file=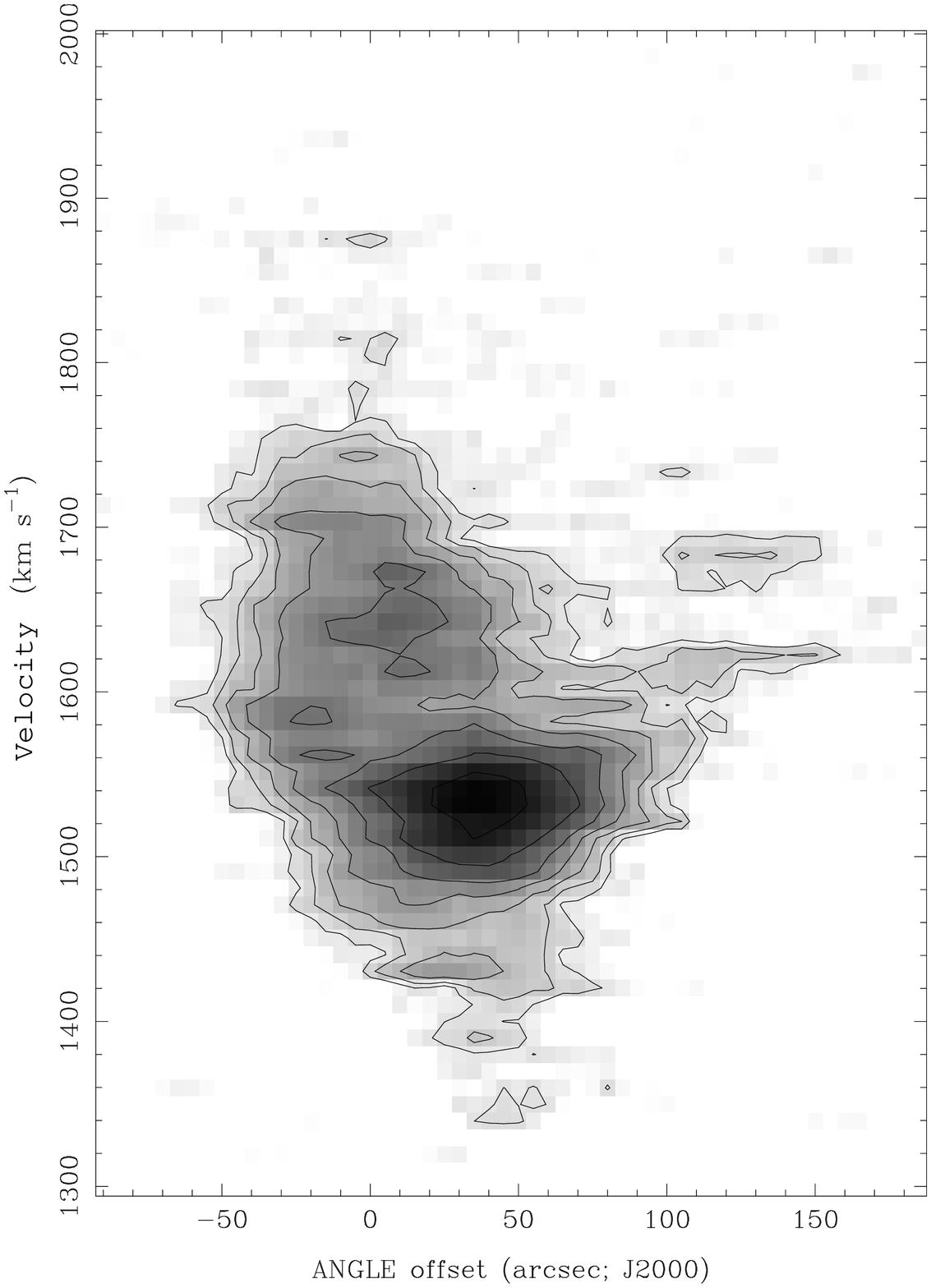,height=9.8cm}}&
\mbox{\psfig{file=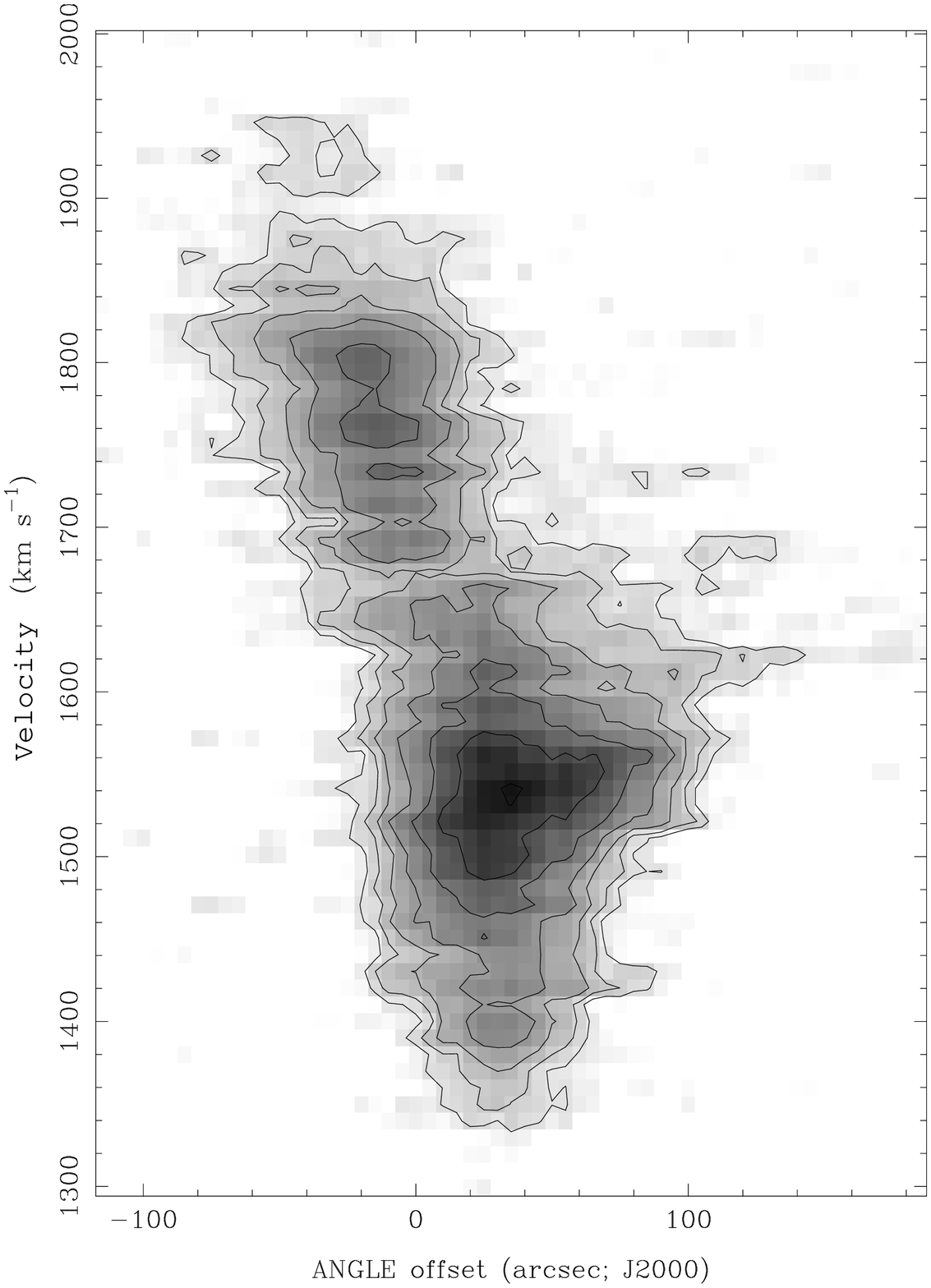,height=9.8cm}}\\
{\bf(a)} Slice along the major axis of NGC~4038 &
{\bf(b)} Slice along the major axis of NGC~4039
\end{tabular}
\caption[\HI\ \pv\ Slices for the Antennae]{
\label{figure:HIPV_majaxis_slice}
\HI\ \pv\ slices, obtained by sampling the channel maps along
lines through the nuclei of each galaxy, roughly along their optical major
axes. East is to the left, west to the right:
{~~\bf(a)~} for the northern disk (NGC~4038), ${\rm PA}\approx98.7\degrE$,
{~~\bf(b)~} for the southern disk (NGC~4039), ${\rm PA}\approx56.5\degrE$.
Contours: $3\sigma\times2^{n/2}\approx2.7$, 3.82, 5.4, 7.64, 10.8, 15.3,
$\baseexpE{21.6}{-3}\,\JypbE$ ($n\geq0$).  }
\end{figure*}


\begin{figure*} 
\begin{center}\begin{tabular}{cc}
\mbox{\psfig{file=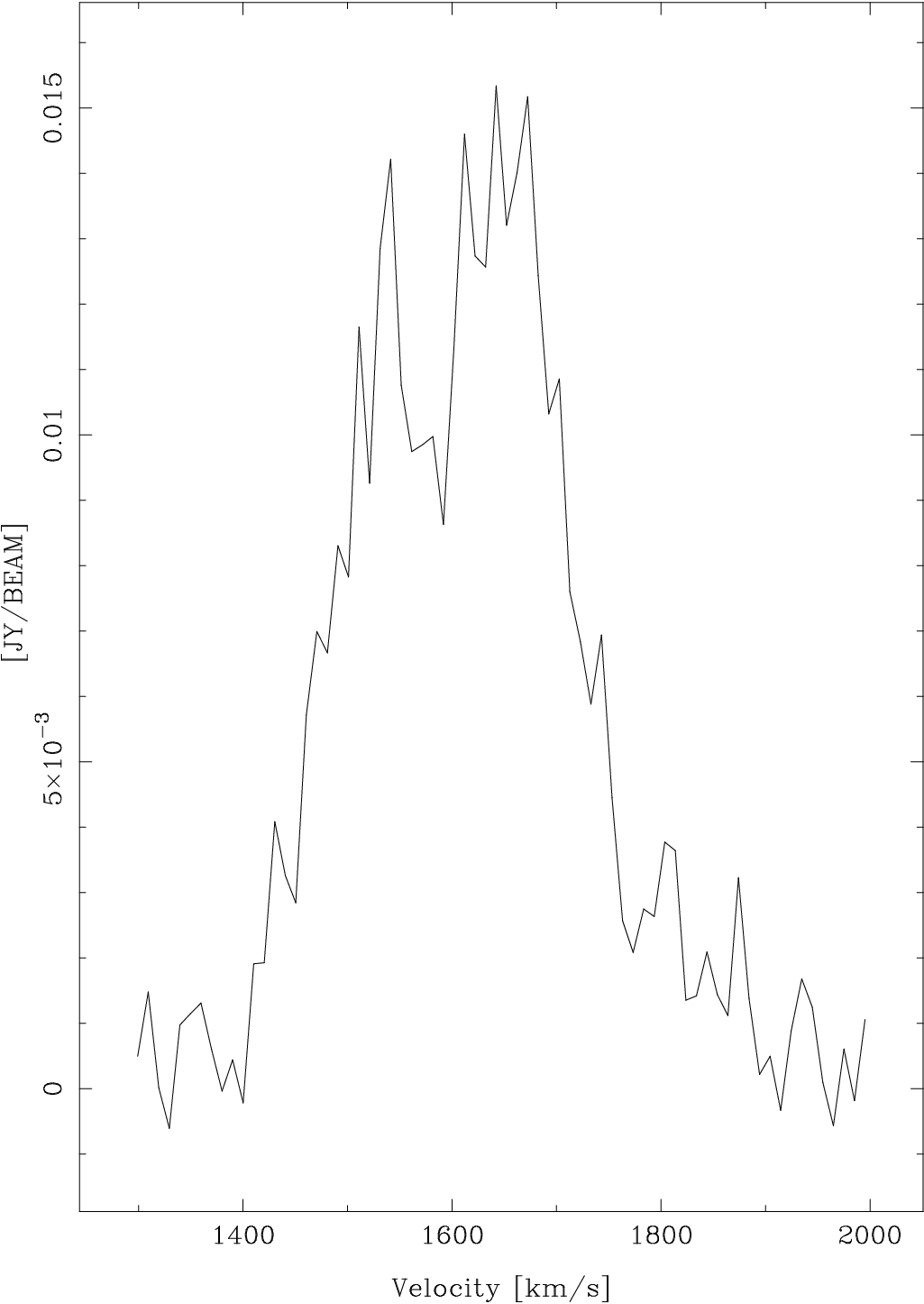,height=9cm}}&
\mbox{\psfig{file=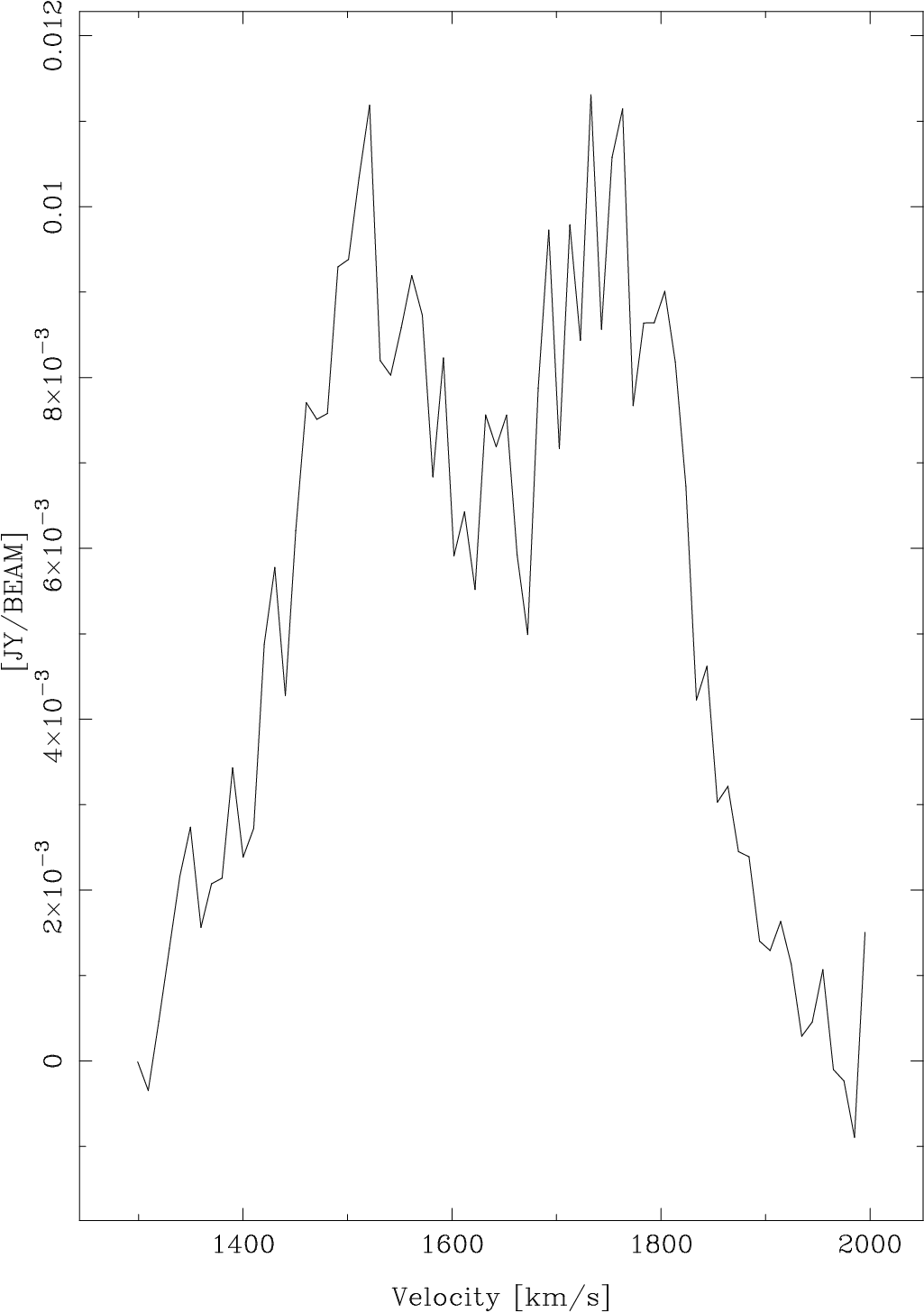,height=9cm}}\\
\end{tabular}\end{center}
\caption[\HI\ Spectra at the NGC~4038/9 Nuclei]{
\label{figure:HINuclearSpectra}
\HI\ spectra at the positions of the two nuclei: {\bf (a)} NGC~4038 (northern
nucleus) and {\bf (b)} NGC~4039 (southern nucleus); see 
Table~\ref{table:TheGroup}). Both spectra show two emission peaks. 
Conversion:
$1~\JypbE\approx\baseexpE{6.23}{20}\,\pcmcmE\pkmpsE\approx342{\rm~K}$.  }
\end{figure*}


\begin{figure*} 
\begin{tabular}{cc}
\mbox{\psfig{file=figure6a.ps,height=10.5cm}} &
\mbox{\psfig{file=figure6b.ps,height=10.5cm}} \\
{\bf(a)} \HI\ spectrum of the Antennae disks &
{\bf(b)} \HI\ spectrum of ESO~572--G045 
\vspace{5mm} \\
\mbox{\psfig{file=figure6c.ps,height=10.5cm}} &
\mbox{\psfig{file=figure6d.ps,height=10.5cm}} \\
{\bf(c)} \HI\ spectrum of the northern tail &
{\bf(d)} \HI\ spectrum of the southern tail
\vspace{5mm}
\end{tabular}
\caption[\HI\ Spectra of the Antennae]{
\label{figure:HISpectra}
\HI\ spectra of various regions in the Antennae group, integrated spatially. 
Conversion: $1~\JyE\approx\baseexpE{8.68}{7}~\MsolarE\pkmpsE$. }
\end{figure*}


\begin{figure*} 
\begin{tabular}{c}
\mbox{\psfig{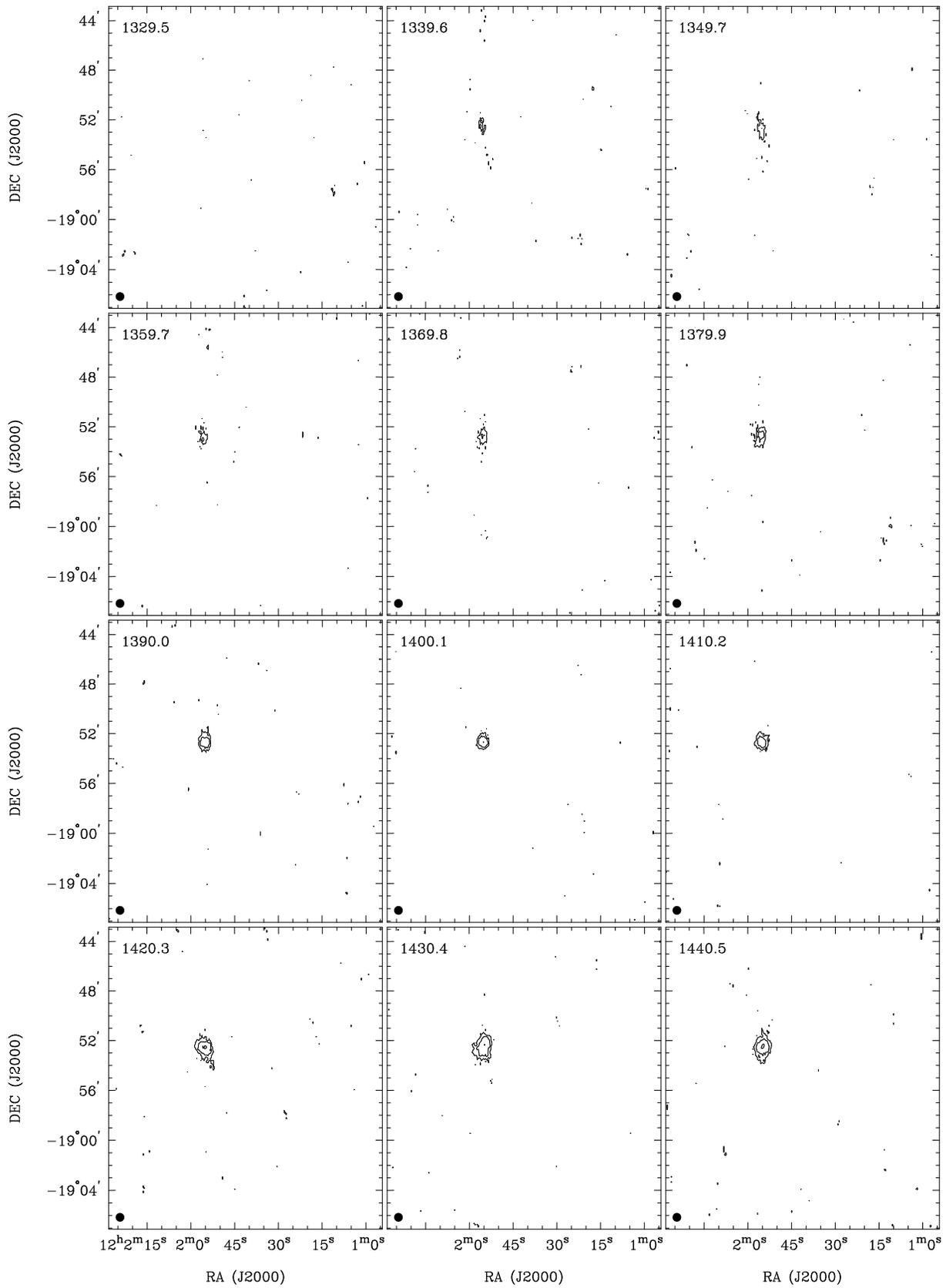}}\\
\end{tabular}
\caption[\HI\ Channel Maps for the Antennae Group]{
\label{figure:HIChannelMaps}
\HI\ channel maps for the Antennae group. Contours: $-3\sigma,~ 3\sigma\times
2^n = -2.7, 2.7, 5.4, 10.8, 21.6, 33.2, 66.4~\mJypbE$, or
$\sim\baseexpE{1.7}{18}\times2^n\,\psqcmE\pkmpsE$ ($n\geq0$). Also
$1~\JypbE\approx342{\rm~K}$ (from Eqn.(12.2) in Giovanelli \& Haynes 1988).
The 42\arcsec\ beam is shown at bottom left.  }
\end{figure*}
\begin{figure*} 
\begin{tabular}{c}
\mbox{\psfig{file=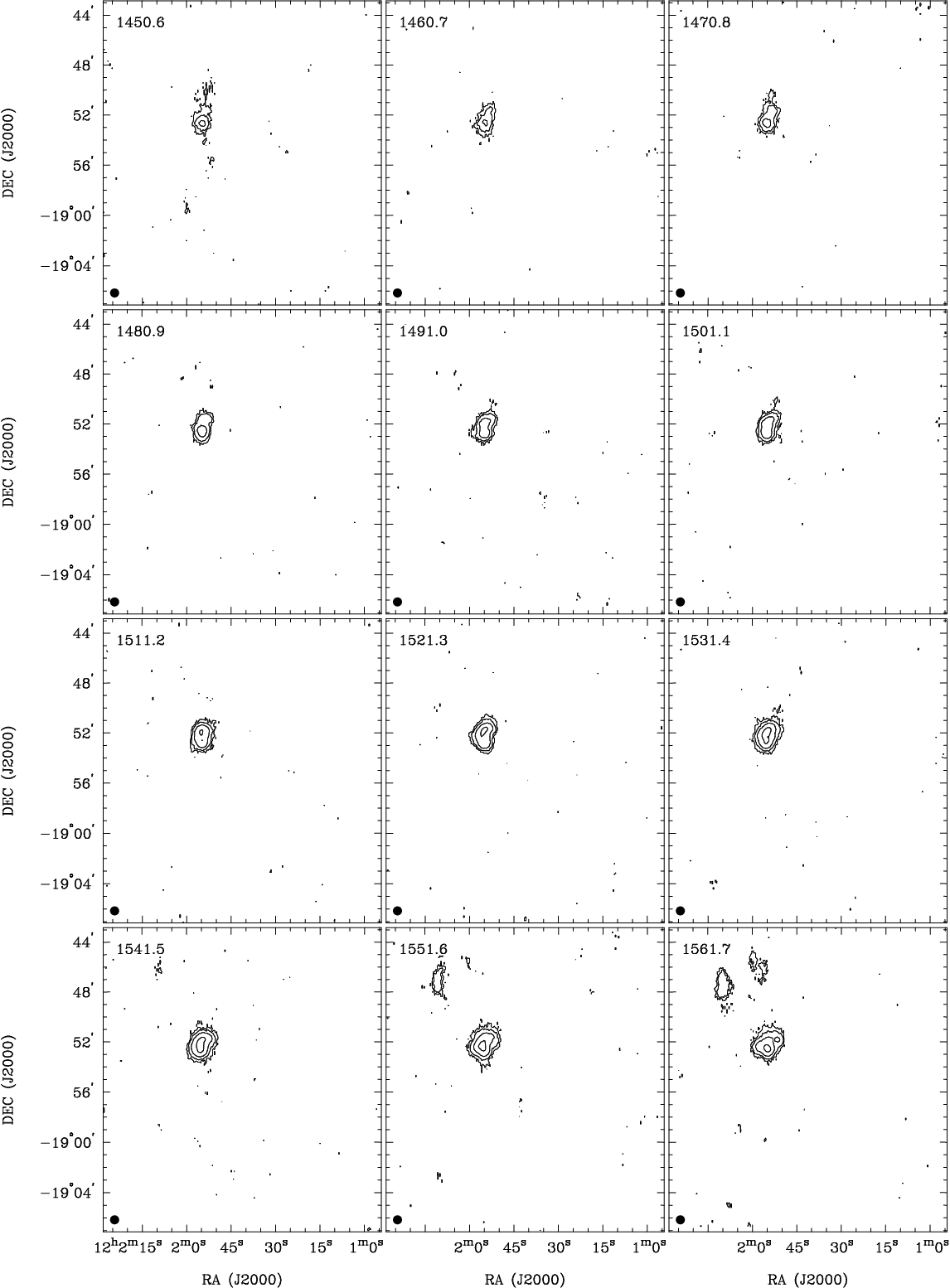,height=22.1cm}}\\
{\bf Figure~\ref{figure:HIChannelMaps}}~~~(continued)
\end{tabular}
\end{figure*}
\begin{figure*} 
\begin{tabular}{c}
\mbox{\psfig{file=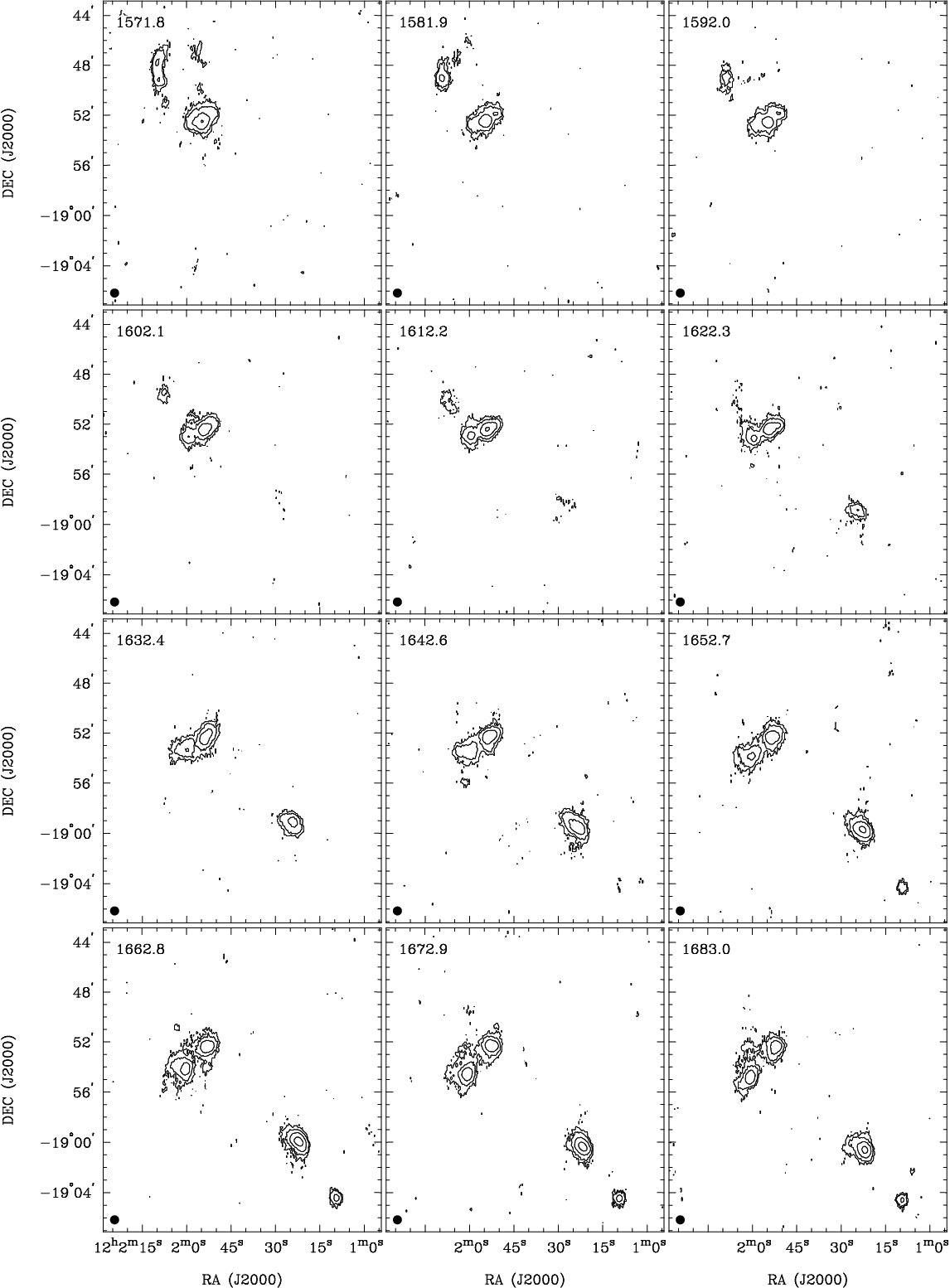,height=22.9cm}}\\
{\bf Figure~\ref{figure:HIChannelMaps}}~~~ (continued)
\end{tabular}
\end{figure*}
\begin{figure*} 
\begin{tabular}{c}
\mbox{\psfig{file=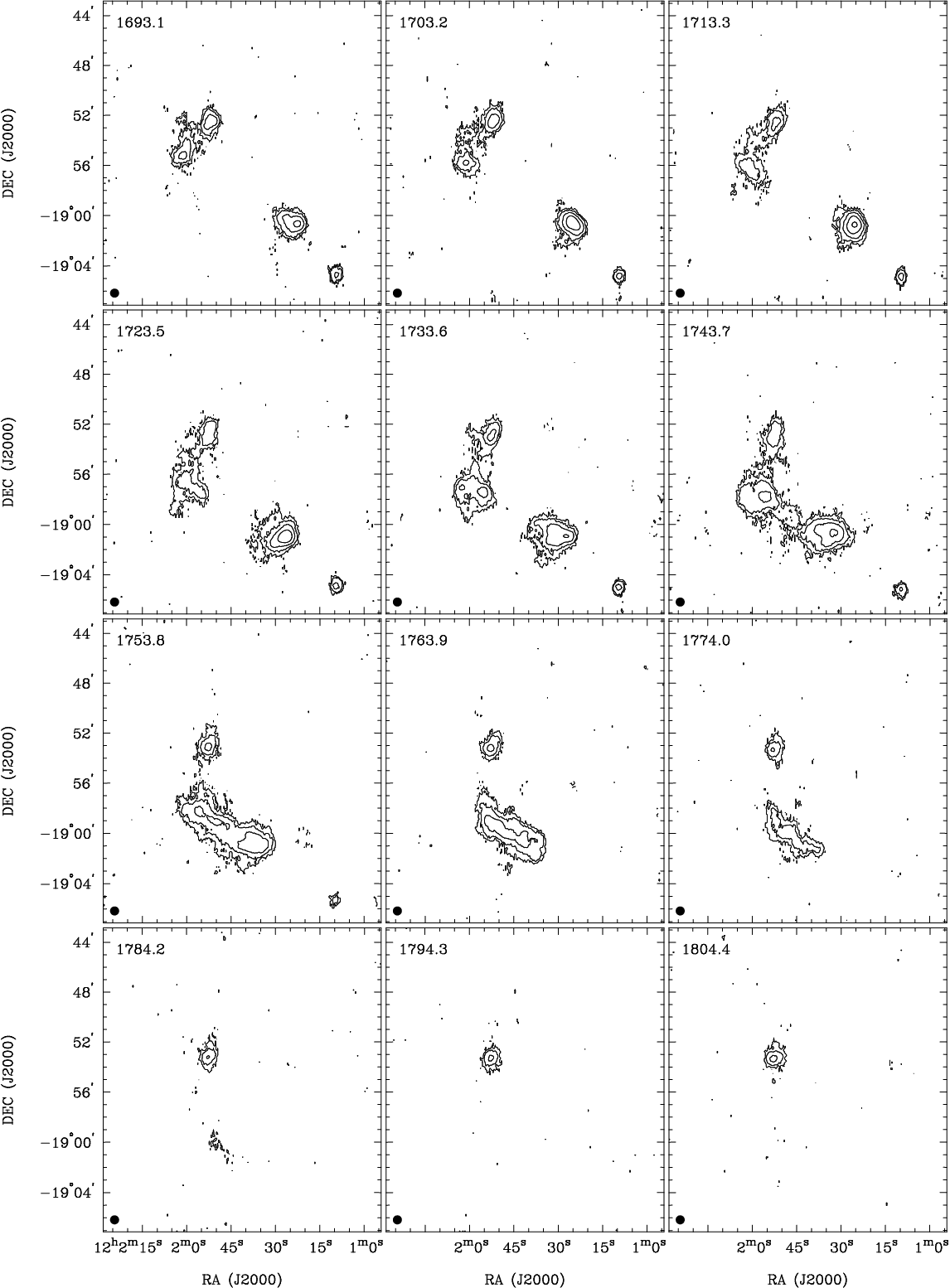,height=22.1cm}}\\
{\bf Figure~\ref{figure:HIChannelMaps}}~~~(continued)
\end{tabular}
\end{figure*}
\begin{figure*} 
\begin{tabular}{c}
\mbox{\psfig{file=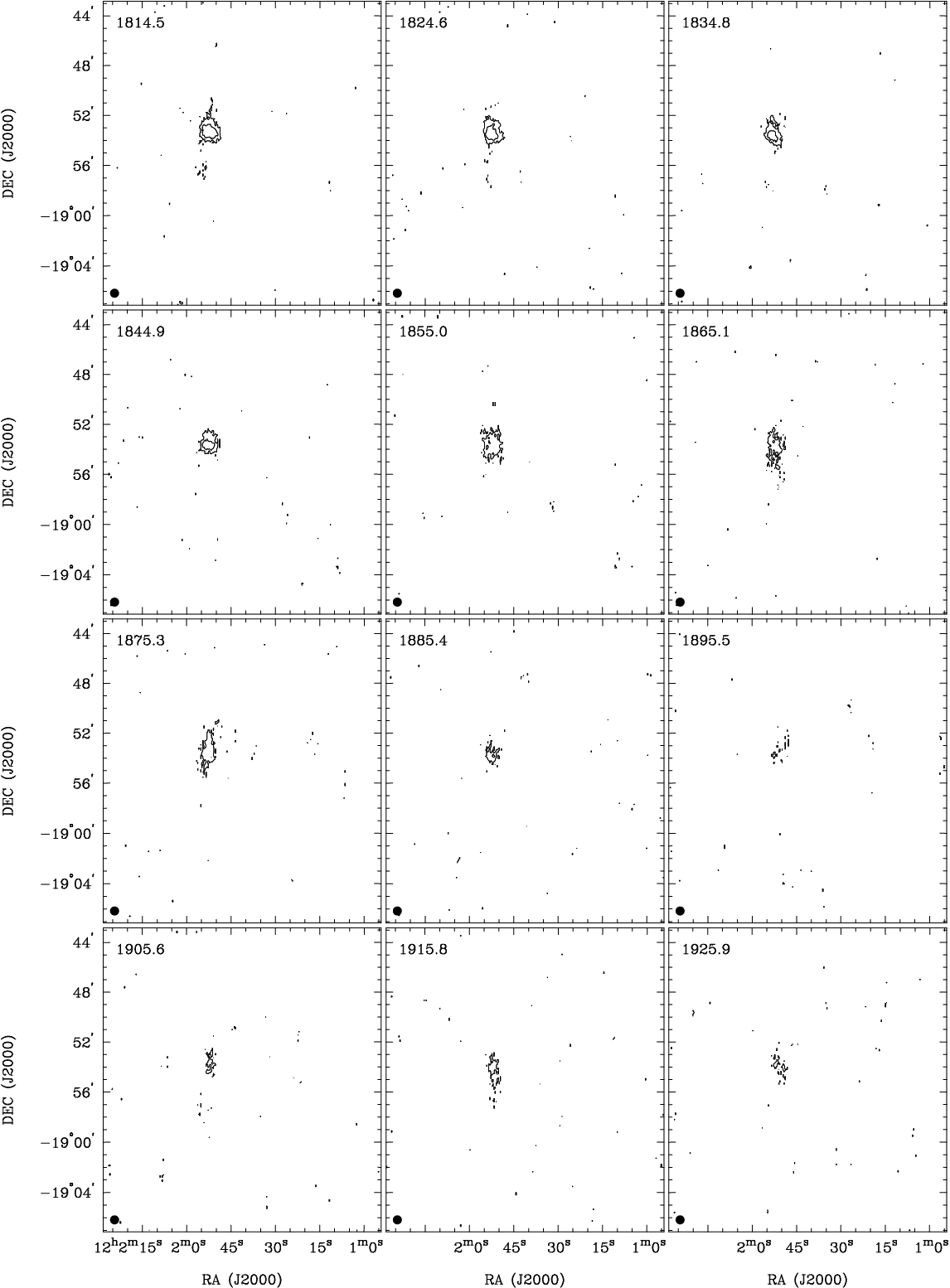,height=22.1cm}}\\
{\bf Figure~\ref{figure:HIChannelMaps}}~~~(continued)
\end{tabular}
\end{figure*}
\begin{figure*} 
\begin{tabular}{c}
\mbox{\psfig{file=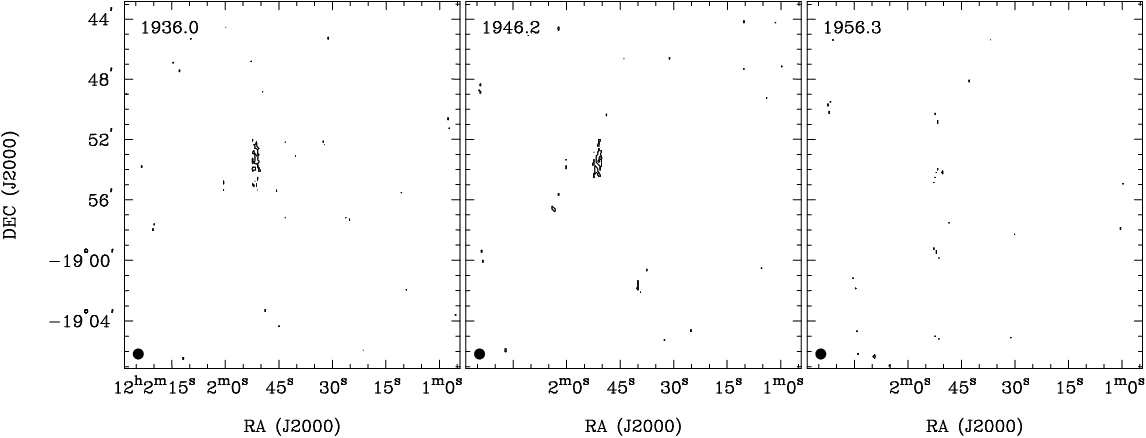,height=6.2cm}}\\
{\bf Figure~\ref{figure:HIChannelMaps}}~~~(continued)
\end{tabular}
\end{figure*}


\begin{figure*} 
\begin{tabular}{cc}
\mbox{\psfig{file=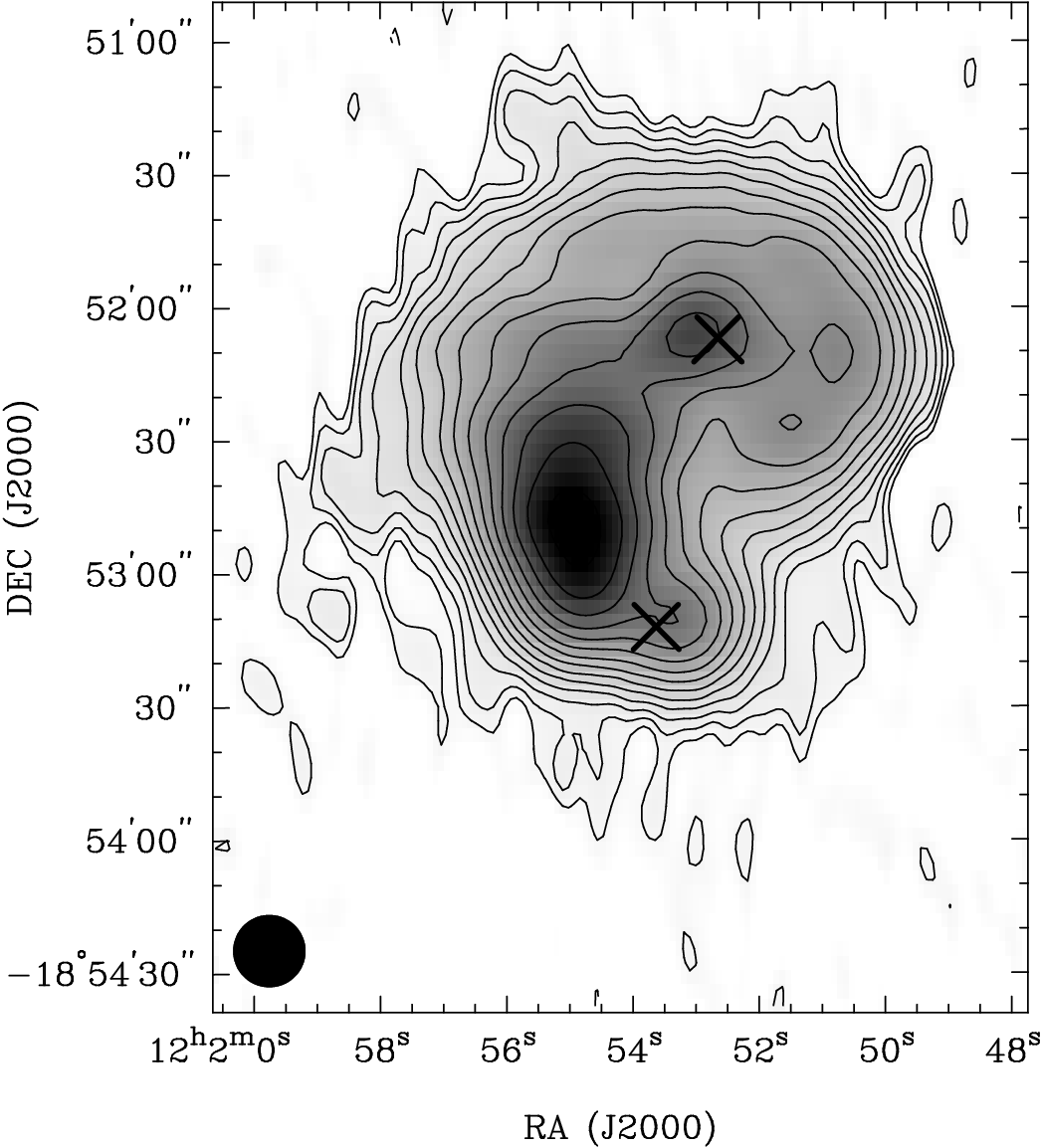,width=8.4cm}} &
\mbox{\psfig{file=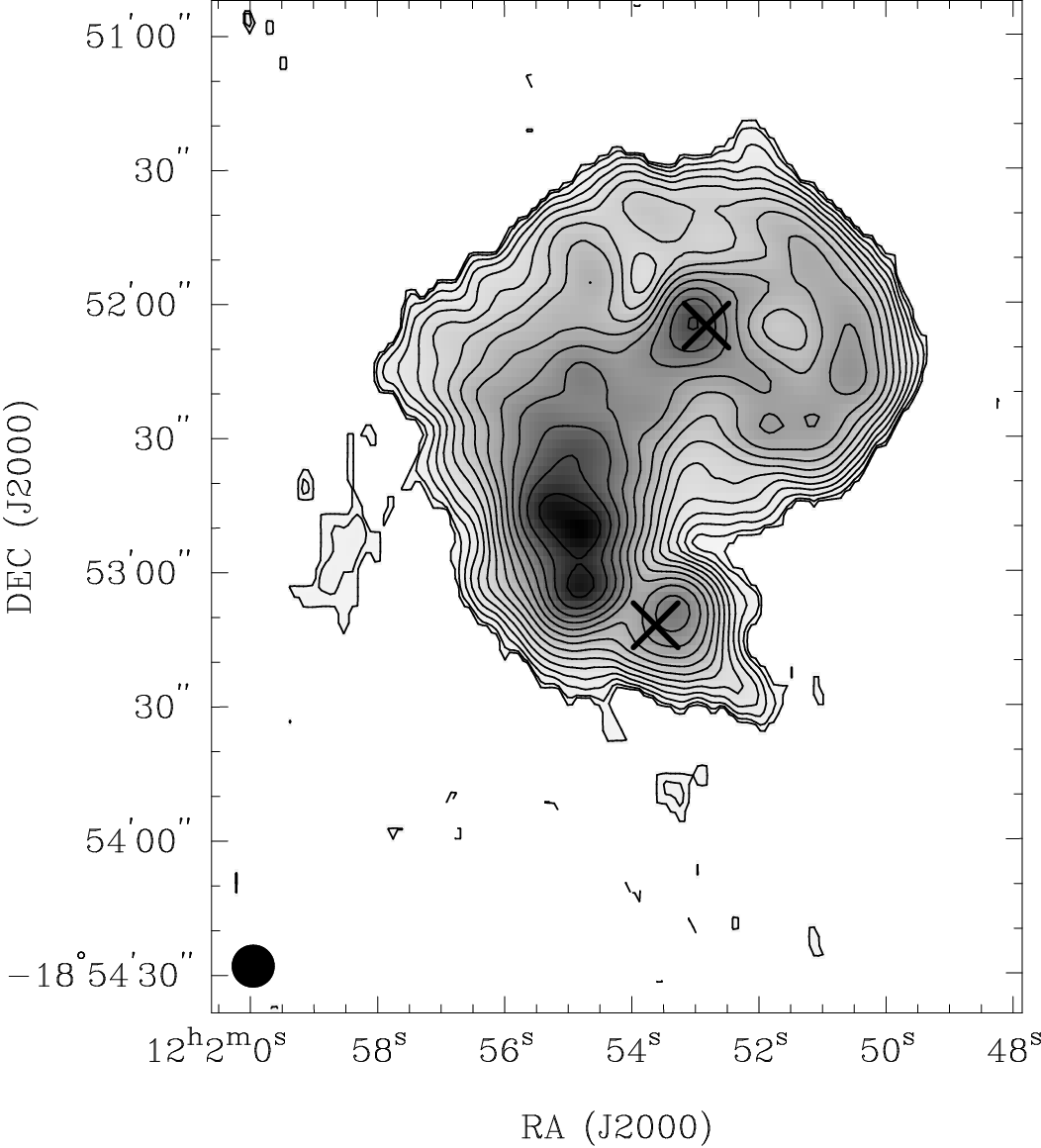,width=8.4cm}} \\
{\bf (a)} 20\,\cm\ radio continuum emission & 
{\bf (b)} 13\,\cm\ radio ontinuum emission
\end{tabular}
\caption[20 and 13\,\cm\ Continuum Maps of the Antennae]{
\label{figure:Antennae20_13CM_Overlay}
Displayed are 20\,\cm\ (left) and 13\,\cm\ (right) radio continuum images of
the Antennae. The beams, shown in the bottom left corner, are 16\arcsec\ and
9\farcs5, respectively, using `uniform' weighting. ~~~~{\bf(a)} 20\,\cm\ 
map: r.m.s. noise $\baseexpE{2.4}{-4}~\JypbE$; contours --0.72, 0.72, 1.02, 
1.44, 2.04, 2.88, 4.07, 5.76, 8.15, 11.5, 16.3, 23.0, 32.6, 46.1, 65.2~\mJypb. 
$1~\mJypbE\approx0.39~\mJypkpckpcE$. ~~~~{\bf(b)} 13\,\cm\ map: r.m.s. noise
$\baseexprangeE{6.1}{6.5}{-5}~\JypbE$; contours --0.18, 0.18, 0.255, 0.360,
0.509, 0.720, 1.02, 1.44, 2.04, 2.88, 4.07, 5.76, 8.15, 11.5, 16.3~\mJypb.
$1~\mJypbE\approx1.12~\mJypkpckpcE$. }
\end{figure*}


\begin{thebibliography}{}

\bibitem{amram1}
Amram P., Marcelin M., Boulesteix J., le Coarer E., 1992, A\&A, 266, 106

\bibitem{arp1}
Arp H.C., 1966, `Atlas of Peculiar Galaxies', California Institute of Technology
(equivalent to Arp H.C., 1966, ApJS, 14, 1)

\bibitem{arp2}
Arp H.C., 1969, A\&A, 3, 418

\bibitem{barnes1}
Barnes J.E., 1988, ApJ, 331, 669

\bibitem{barnes2}
Barnes J.E., Hernquist L., 1992, Nature, 360, 715

\bibitem{barnes3}
Barnes J.E., Hernquist L., 1996, ApJ, 471, 115

\bibitem{clark1}
Clark B.G., 1980, A\&A, 89, 377

\bibitem{clutton1}
Clutton-Brock M., 1972, Ap\&SS, 17, 292

\bibitem{condon1}
Condon J.J., 1983, ApJS, 53, 459

\bibitem{condon2}
Condon J.J., 1987, ApJS, 65, 485

\bibitem{devaucouleurs1}
de Vaucouleurs G., de Vaucouleurs A., Corwin H.G. Jr. , Buta R.J.,
  Paturel G., Fouqu\'e P., 1991, `Third Reference Catalogue of Bright
  Galaxies', New York: Springer Verlag [RC3]

\bibitem{douglas1}
Douglas J.N., Bash F.N., Bozyan F.A., Torrence G.W., Wolfe C., 1996, AJ, 111, 1945

\bibitem{duc1}
Duc P.-A., Mirabel I.F., 1994, A\&A, 289, 83

\bibitem{duc2}
Duc P.-A., Mirabel I.F., 1998, A\&A, 333, 813

\bibitem{elmegreen1}
Elmegreen D.M., Kaufman M., Brinks E., Elmegreen B.G., Sundin M., 1995,
ApJ, 453, 100

\bibitem{elmegreen2}
Elmegreen B.G., Kaufman M., Thomasson M., 1993, ApJ, 412, 90 

\bibitem{elmegreen3}
Elmegreen B.G., Sundin M., Kaufman M., Brinks E., Elmegreen B.G., 1995, ApJ, 453, 139

\bibitem{fabbiano1}
Fabbiano G., Schweizer F., Mackie G., 1997, ApJ, 478, 542

\bibitem{fabbiano2}
Fabbiano G., Trinchieri G., 1983, ApJ, 266, L5

\bibitem{fabbiano3}
Fabbiano G., Zezas A., Murray S.S., 2000, BAAS, 32, 1532
[Meeting 197, abstract 79.01]

\bibitem{fischer1}
Fischer J., \etal\, 1996, A\&A, 315, L97

\bibitem{gao1}
Gao Y., Gruendl R.A., Lo K.Y., Lee S.-W., Hwang C.Y., 1998, BAAS, 30, 923
[Meeting 192, abstract 69.04]

\bibitem{gao2}
Gao Y., Lo K.Y., Lee S.-W., Lee T.-H, 2001, ApJ, 548, 172

\bibitem{gioia1}
Gioia I.M., Maccacaro T., Schild R.E., Wolter A., Stocke J.T., Morris S.L.,
  Henry J.P., 1990, ApJS, 72, 567

\bibitem{giovanelli1}
Giovanelli R., Haynes M.P., 1988, in `Galactic and Extragalactic Radio 
  Astronomy', Verschuur, G.L. \& Kellerman, K.I. (eds), Springer-Verlag, 
  p.\,522

\bibitem{hibbard1}
Hibbard J.E., 1995, PhD Thesis (Columbia University)

\bibitem{hibbard2}
Hibbard J.E., Mihos J.C., 1995, AJ, 110, 140

\bibitem{hibbard3}
Hibbard J.E., Vacca W.D., Yun, M.S., 2000, AJ, 119, 1130

\bibitem{hibbard4}
Hibbard J.E., van der Hulst J.M., Barnes J.E., 2001, in prep.

\bibitem{hibbard5}
Hibbard J.E., van Gorkom J.H., 1996, AJ, 111, 655

\bibitem{hogbom1}
H\"ogbom J.A., 1974, A\&AS, 15, 417

\bibitem{huchtmeier1}
Huchtmeier W.K., Bohnenstengel H.D., 1975, A\&A, 41, 477

\bibitem{hummel1}
Hummel E., van der Hulst J.M., 1986, A\&A, 155, 151

\bibitem{karachentsev1}
Karachentsev I.D., Karachentseva V.E., Parnovskij S.L., 1993, Astronomische
  Nachrichten, 314, 97

\bibitem{kaufman1}
Kaufman M., Brinks E., Elmegreen B.G., Elmegreen D.M., Struck C., Thomasson M.,
  Klaric M., 1995, BAAS, 186, \#39.10

\bibitem{kunze1}
Kunze D., \etal, 1996, A\&A, 315, L101

\bibitem{maccacaro1}
Maccacaro T., della Ceca R., Gioia I.M., Morris S.L., Stocke J.T., Wolter A.,
  1991, ApJ, 374, 117

\bibitem{mahoney1}
Mahoney J.H., Burke B.F., van der Hulst J.M., 1987, IAU Symposium No. 117, 94

\bibitem{malphrus1}
Malphrus B.K., Simpson C.E., Gottesman S.T., Hawarden T.G., 1997, ApJ, 114, 1427

\bibitem{mihos1}
Mihos J.C., Bothun G.D., Richstone D.O., 1993, ApJ, 418, 82

\bibitem{mihos2}
Mihos J.C., Hernquist L., 1994, ApJ, 431, L9

\bibitem{mihos3}
Mihos J.C., Richstone D.O., Bothun G.D., 1991, ApJ, 377, 72

\bibitem{minkowski1}
Minkowski R., 1957, IAU Symposium No. 4, 107, Cambridge University Press

\bibitem{mirabel1}
Mirabel I.F., Dottori H., Lutz D., 1992, A\&A, 256, L19

\bibitem{mirabel2}
Mirabel I.F., Lutz D., Maza J., 1991, A\&A, 243, 367

\bibitem{mirabel3}
Mirabel I.F., Vigroux L., \etal\, 1998, A\&A, 333, L1

\bibitem{neff1}
Neff S.G., Ulvestad J.S. 2000, AJ, 120, 670

\bibitem{nikola1}
Nikola T., Genzel R., Herrmann F., Madden S.C., Poglitsch A., Geis N.,
Townes C.H., Stacey G.J., 1998, ApJ, 504, 749

\bibitem{parker1}
Parker P.R., 1990, `Colliding galaxies: the universe in turmoil', 
  Plenum Press, p.\,171

\bibitem{pisano1}
Pisano D.J., Wilcots E.M., 2000, MNRAS, 319, 821

\bibitem{read1}
Read A.M., Ponman T.J., Wolstencroft R.D., 1995, MNRAS, 277, 397

\bibitem{roberts1}
Roberts M.S., 1975, in `Galaxies and the Universe', Sandage, A., Sandage, M.
  \& Kristian, J. (eds), Chicago: University of Chicago Press, p.\,309

\bibitem{rubin1}
Rubin V.C., Ford W.K., D'Odorico S., 1970, ApJ, 160, 801

\bibitem{sanders1}
Sanders D.B., Mirabel I.F., 1985, ApJ, 298, L31

\bibitem{sanders2}
Sanders D.B., Mirabel I.F., 1996, ARA\&A, 34, 749

\bibitem{sansom1}
Sansom A.E., Dotani T., Okada K., Yamashita A., Fabbiano G., 1996,
MNRAS, 281, 48

\bibitem{sault1}
Sault R.J., Teuben P.J., Wright M.C.H., 1995, in `Astronomical Data 
  Analysis Software and Systems IV', ASP Conf.Series 77, p.\,433

\bibitem{schweizer1}
Schweizer F., 1978, in `Structure and Properties of Nearby Galaxies', 
  Berkhuijsen, E.M. \& Wielebinski, R. (eds), IAU Symposium No. 77, p.\,279

\bibitem{shapley1}
Shapley H., Paraskevopoulos J.S., 1940, Proc. Nat. Acad. Sci., 26, 31

\bibitem{stanford1}
Stanford S.A., Sargent A.I., Sanders D.B., Scoville N.Z., 1990, ApJ, 349, 492

\bibitem{steer1}
Steer D.G., Dewdney P.E., Ito M.R., 1984, A\&A, 137, 159

%
%
\bibitem{stocke1}
Stocke J.T., \etal\ 1991, ApJS, 76, 813

\bibitem{toomre1}
Toomre A., 1977, in `Evolution of Galaxies and Stellar Populations',
  Tinsky, B.M. \& Larson, R.B. (eds), Yale University Observatory, 
  New Haven, p.\,401

\bibitem{toomre2}
Toomre A., Toomre J., 1972, ApJ, 178, 623

\bibitem{vanderhulst1}
van der Hulst J.M., 1979, A\&A, 71, 131

\bibitem{vigroux1}
Vigroux L., Mirabel F., \etal, 1996, A\&A, 315, L93

\bibitem{whitmore1}
Whitmore B.C., Schweizer F., 1995, AJ, 109, 960

\bibitem{whitmore4}
Whitmore B.C., Zhang Q., Leitherer C., Fall S.M., Schweizer F., Miller B.W.,
1999, AJ, 118, 1551

%
\bibitem{wilson1}
Wilson C.D., Scoville N., Madden S.C., Charmandaris V., 2000, ApJ, 542, 120

\bibitem{}
Young J.S., \etal, 1995, ApJS, 98, 219

\end{thebibliography}
\end{document}